\documentclass[
  journal=pasa,
  manuscript=research-paper, 
  year=2024,
  volume=37,
  useAMS
]{cup-journal}

\usepackage[nopatch]{microtype}
\usepackage{booktabs}
\usepackage{graphicx}	
\usepackage{subfig}
\usepackage{microtype,siunitx,booktabs,amssymb,amsmath,bm}
\def\code#1{\texttt{#1}}

\title{Optimising the MeerKAT Pulsar Timing Array and towards precision pulsar timing with SKA-mid}

\author{Pratyasha Gitika}
\affiliation{Centre for Astrophysics and Supercomputing, Swinburne University of Technology, Mail H39, PO Box 218, VIC 3122, Australia.}
\alsoaffiliation{ARC Centre of Excellence for GW Discovery (OzGrav), Swinburne University of Technology, Mail H11, PO Box 218, VIC 3122.}
\email[Pratyasha Gitika]{pgitika@swin.edu.au}

\author{Ryan M. Shannon}
\affiliation{Centre for Astrophysics and Supercomputing, Swinburne University of Technology, Mail H39, PO Box 218, VIC 3122, Australia.}
\alsoaffiliation{ARC Centre of Excellence for GW Discovery (OzGrav), Swinburne University of Technology, Mail H11, PO Box 218, VIC 3122.}

\author{Matthew Bailes}
\affiliation{Centre for Astrophysics and Supercomputing, Swinburne University of Technology, Mail H39, PO Box 218, VIC 3122, Australia.}
\alsoaffiliation{ARC Centre of Excellence for GW Discovery (OzGrav), Swinburne University of Technology, Mail H11, PO Box 218, VIC 3122.}

\author{Daniel J. Reardon}
\affiliation{Centre for Astrophysics and Supercomputing, Swinburne University of Technology, Mail H39, PO Box 218, VIC 3122, Australia.}
\alsoaffiliation{ARC Centre of Excellence for GW Discovery (OzGrav), Swinburne University of Technology, Mail H11, PO Box 218, VIC 3122.}

\author{Matthew T. Miles}
\affiliation{Centre for Astrophysics and Supercomputing, Swinburne University of Technology, Mail H39, PO Box 218, VIC 3122, Australia.}
\alsoaffiliation{ARC Centre of Excellence for GW Discovery (OzGrav), Swinburne University of Technology, Mail H11, PO Box 218, VIC 3122.}

\author{David J. Champion}
\affiliation{Max-Planck-Institut für Radioastronomie, Auf dem Hügel 69, D-53121 Bonn, Germany.}

\author{Kathrin Grunthal}
\affiliation{Max-Planck-Institut für Radioastronomie, Auf dem Hügel 69, D-53121 Bonn, Germany.}

\keywords{millisecond pulsars, gravitational waves, radio astronomy, interstellar medium} 

\begin{document}

\begin{abstract}
Pulsar timing arrays (PTAs) are Galactic-scale nanohertz-frequency gravitational wave (GW) detectors. Recently, several PTAs
have found evidence for the presence of GWs in their datasets, but none of them have achieved a community-defined definitive (> 5$\sigma$)
detection. Here, we identify limiting noise sources for PTAs and quantify their impact on sensitivity to GWs
under different observing and noise modelling strategies. First, we search for intrinsic pulse jitter in a sample of 89 millisecond pulsars (MSPs) observed by the MeerKAT Pulsar Timing Array and obtain new jitter measurements for 20 MSPs. We then forecast jitter
noise in pulsars for the future SKA-Mid telescope, finding that the timing precision of many of the best-timed
MSPs would be dominated by jitter noise. We then consider dispersion measure variations from the interstellar medium and find that their effects are best mitigated by modelling them as a
stationary Gaussian process with a power-law spectrum. Improving upon the established \code{hasasia} code for PTA sensitivity analysis, we assess the timing potential of the lower frequency UHF-band (544$-$1088\,MHz) of MeerKAT and find a potential increase
in GW background sensitivity by $\approx 8$\%, relative to observing at L-band. 
We show that this improvement relies on assumptions on the propagation through the interstellar medium, and highlight that if observing frequency-dependent propagation effects, such as scattering noise, are present, where noise is not completely correlated across observing frequency, then the improvement is significantly diminished.
Using the multi-frequency receivers and sub-arraying
flexibility of MeerKAT, we find that focussed, high-cadence observations of the best
MSPs can enhance the sensitivity of the array for both the continuous GWs and
stochastic GW background. These results highlight the role of MeerKAT and the MPTA in the context of international GW search
efforts.

\end{abstract}

\section{Introduction}

Pulsar timing arrays (PTAs) aim to detect nanohertz (nHz) frequency gravitational waves (GW) by monitoring radio pulse times of arrivals (ToAs) from an ensemble of millisecond pulsars (MSPs) \citep{1982Natur.300..615B, 1990ApJ...361..300F}.
The signature of nHz GWs can be detected by analysing the spatial correlations in the timing residuals, i.e. the difference between the predicted and observed ToAs.
 The population of supermassive black hole binaries (SMBHBs) in the universe is the most plausible source of GWs in the PTA band \citep{2003ApJ...583..616J, 2003ApJ...595..614W}. 
Even though other potential sources can emit nHz frequency GWs \citep{2003ApJ...583..616J, 2005PhyU...48.1235G, 2010PhRvD..81j4028O, 2023MNRAS.521.5077F}, the first detection by PTAs is expected to come from an incoherent superposition of GWs from an unresolved
population of SMBHBs, known as the stochastic GW background
(SGWB). The expected inter-pulsar correlation of the timing residuals
as a function of the angular separation of the Earth-pulsar baselines for an SGWB is known as the Hellings-Downs (HD)
curve \citep{1983ApJ...265L..39H, 2024CQGra..41q5008R}, and arises due to the quadrupolar nature of the distortions of space-time induced by gravitational waves.
Detecting a quadrupolarly correlated signal is necessary to confirm the presence of nHz-frequency GWs.

Recently, the Chinese pulsar timing array \citep[CPTA,][]{2023RAA....23g5024X}, European pulsar timing array (EPTA) and Indian Pulsar timing array (InPTA) \citep{2023A&A...678A..50E},  the North American Nanohertz Observatory for Gravitational Waves  \citep[NANOGrav,][]{2023ApJ...951L...8A}, and the Parkes Pulsar Timing Array  \citep[PPTA,][]{2023ApJ...951L...6R}, presented searches of GWs in their most recent datasets.  
Even though the individual PTA data analysis and pulsar samples are different, they all identified a
common spatially-uncorrelated red noise (common to all pulsars without the HD correlation; CURN) with evidence ($\sim$2$-$4$\sigma$) for an HD-correlated SGWB. 
More recently, the MeerKAT pulsar timing array (MPTA) has studied the largest sample of the Southern hemisphere pulsars and found evidence (3.4$\sigma$) for the presence of an SGWB \citep{2025MNRAS.536.1489M}. 
These established and emerging PTAs form the International Pulsar timing array  \citep[IPTA,][]{2010CQGra..27h4013H}.
Even though the PTAs have found evidence for an SGWB, none of the results have reached a statistical significance equivalent or above 5$\sigma$, the threshold established by the IPTA to ensure a robust detection \citep{2023arXiv230404767A}. 
To increase the significance of detection and characterise the source of any purported GW signal, PTAs should fine-tune their `detector' for enhanced sensitivity \cite[][]{2013CQGra..30v4015S}. 

The sensitivity of a PTA depends on the characteristics of the GW signal being observed. For the detection of an SGWB, optimal sensitivity can be achieved by observing more pulsars for a long period of time, which increases the number of pulsar pairs effectively contributing towards HD-correlations. On the other hand, for a single source, careful time allocation towards a few best pulsars is potentially more optimal \citep{2011ApJ...730...17B}. In practice, observing is constrained by the limits to observing time.  Previous studies have explored sensitivity changes by careful allocation of each pulsar observation time, removing poorly timed pulsars, using longer or high-cadence observations, or improving the significance of detection towards multiple targets of continuous GWs \citep{2012MNRAS.423.2642L, 2018ApJ...868...33L, 2021CQGra..38e5009K, 2023MNRAS.518.1802S}.
More recently, \cite{2024arXiv240900336B} simulated an IPTA-like array and tested various observing campaigns and sensitivities. Their study concluded that combining high-cadence observations of precisely timed pulsars and low-cadence observations of the less sensitive pulsars optimised a PTA towards both single sources and SGWB.

The simplest way to improve a PTA is by increasing ToA precision through higher signal-to-noise ratio (S/N) observations using more sensitive telescopes. However, pulsar timing residuals show excess scatter beyond that estimated from the thermal (radiometer) noise of the receiver. This scatter is usually modelled as a time-uncorrelated white noise component or a time-correlated red noise component. One such white noise component can arise from intrinsic variations in the pulse morphology of a pulsar, which is termed jitter noise \citep{1975ApJ...198..661H, 1998ApJ...498..365J,2022MNRAS.510.5908M}. 
The ToAs are derived from cross-correlating a model pulse profile, known as a template, with an observed integrated pulse profile, which is the average of a finite number of pulses.
Stochastic variations in pulse shape, phase and intensity cause these integrated pulse profiles to vary, contributing to jitter noise ranging from 4\,ns $\rm hr^{-1}$ to 100\,ns $\rm hr^{-1}$ \citep{2021MNRAS.502..407P}. Jitter noise becomes significant for high S/N observations and dominates the radiometer noise.
While longer integration time can reduce jitter noise, it requires more telescope time, making it essential to characterise for each pulsar. 
Previous studies of timing noise for the PPTA \citep{2014MNRAS.443.1463S}, NANOGrav \citep{2019ApJ...872..193L}, MeerKAT \citep{2021MNRAS.502..407P}, and FAST \citep{2024ApJ...964....6W} MSPs, respectively, conclude that jitter noise dominates the white noise budget for more sensitive observations. Therefore, characterisation of this noise source is important for GW detection.

In pulsar timing, red noise is typically categorised as radio frequency independent (achromatic red noise) or radio frequency dependent (chromatic red noise). 
Potential sources of achromatic red noise are
rotational irregularities \citep{2010MNRAS.402.1027H,2010ApJ...725.1607S} and the influence of an SGWB, emerging as a CURN signal. Frequency-dependent red noise includes dispersion measure (DM) noise due to the variations in the electron column density as the Earth-pulsar line of sight traces different parts of the ionised interstellar medium (IISM) \citep{2016ApJ...817...16C} and scattering or chromatic noise due to variable scattering of radio waves through the inhomogeneous IISM \cite[][]{2010arXiv1010.3785C,2017MNRAS.464.2075S}.
Modelling DM variations accurately is one of the biggest challenges of PTAs. Different approaches have been taken to model DM noise. Most PTAs use a stationary Gaussian process (GP) \citep{2014PhRvD..90j4012V} to model DM noise as a red noise signal, whereas alternative models use a piece-wise constant function, \citep[DMX,][]{2015ApJ...813...65N} to measure DM at each observation epoch comprised of multi-frequency observations. 
Since the recovered SGWB parameters (spectral index and amplitude) can potentially be impacted based on the choice of DM noise modelling \citep{2023ApJ...951L...8A, 2024A&A...692A.170I}, it is important to assess the effect of different models on GW sensitivity.

The MeerKAT radio telescope has emerged as an important instrument for PTA science as it has a low system temperature ($\sim$ 18 K at 1400\,MHz), high gain ($\sim$ 2.8\,K $\rm Jy^{-1}$), and the highest number of MSPs (currently 83) that are regularly observed for GW search \citep{2016mks..confE...1J, 2020PASA...37...28B}. With 64 antennas, MeerKAT also possesses sub-arraying flexibility and can observe multiple targets in different parts of the sky or the same target with multiple receivers simultaneously.
The current MPTA data releases, \cite{2023MNRAS.519.3976M} and \citet{2025MNRAS.536.1467M}, consist of timing observations using the L-band receiving system, which operates at $856$-$1712$\,MHz, as that was the only receiver available for observations initially. 
While other receiving systems operating at higher and lower frequencies are now commissioned, observations were maintained at this frequency to produce a consistent and homogeneous data set and avoid the introduction of additional system or band-dependent noise processes \cite[][]{2016MNRAS.458.2161L}. 
However, in the future, the MPTA could potentially benefit from having a different observing strategy, particularly involving the more recently commissioned Ultra-High Frequency (UHF) receiver spanning 544$-$1088\,MHz and the S-band receiver, which can record 856\,MHz of bandwidth within the range 1750 to 3500\,MHz. As pulsars are usually steep-spectrum objects \citep{1991ApJ...378..687F}, pulsar observations at the UHF-band provide high S/N observations with low ToA uncertainty \citep{2018MNRAS.473.4436J, 2022PASA...39...27S}. Comparatively, MeerKAT's S-band receiver would likely provide lower S/N observations, but frequency-dependent effects such as scattering would be less severe. 

In this work, we investigate potential improvements in the sensitivity through changes in the noise modelling and observing strategy of the MPTA. 
In Section \ref{sec:2}, we describe our data reduction process. Section \ref{sec:3} describes the methodology and measurements of jitter noise. In Section \ref{sec:4} we describe the different optimisation strategies for the MPTA, including the DM noise modelling and observation strategies. We discuss the results in Section \ref{sec:5} and conclude and motivate future work in Section \ref{sec:6}.

\section{Data reduction} \label{sec:2}
\subsection{Dataset}

We base our results on observations of MSPs collected with the MeerKAT radio telescope obtained as part of the MeerTIME Large Survey Project \cite[][]{2020PASA...37...28B}.
The MeerKAT radio telescope comprises 64 13.9-m diameter antennas located in the Karoo region of South Africa. Initially, 189 MSPs were observed as part of a census of the MSP population observable with MeerKAT \cite[][]{2022PASA...39...27S}.  The purpose of the census was to characterise the Galactic field MSPs at L-band and assess the achievable timing precision.
Following the conclusion of the census, a sub-sample of MSPs has been observed regularly. The pulsars and integration times were chosen such that the largest sample of MSPs could be observed to sub-microsecond rms error within $\approx 300$\,h of observing per year. 
The current sample includes $83$ MSPs. Currently, each pulsar is observed at an approximately two-week cadence. A minimum 256\,s is allocated if the required timing precision can be achieved in less time, otherwise, MSPs are observed for a duration of up to $2048$\,s. 
However, there are additional observations in the data set of longer duration, i.e. up to several hours, as the data set also includes observations taken as part of the MeerTIME Relativistic Binary project \citep{2021MNRAS.504.2094K}. 

Nearly all of the observations we consider have been taken with the L-band receiver for recording data at 856$-$1712\,MHz.
We have used the \textit{fold} mode dataset, which folds the digitised, channelised data stream at the topocentric period of the pulsar. The resulting observations have 8\,s sub-integrations with four polarisation states and 1024 phase bins per pulse period, written in \code{PSRFITS} file format. 
To account for bandpass roll-off effects, 48 frequency channels (each with $\sim 0.8\, \rm MHz$ bandwidth) from the top and bottom of the band have been removed. For timing studies, the 928 channels are then averaged to 16 frequency channels with a bandwidth of ~54 MHz each.
As MSPs are faint radio sources, mitigation of radio frequency interference (RFI) is essential. RFI excision is performed using \code{meerguard}, an extension of \code{coastguard} software package \cite[][]{2016MNRAS.458..868L} tailored for  MeerKAT pulsar timing data.

Even though only 83 pulsars are currently monitored as part of the MPTA, we have studied jitter in the larger sample of 89 MSPs that were initially included in the MPTA for long-term monitoring.
To measure jitter noise, we first identified the highest S/N observations taken between 2019 February and 2023 June because these would be the most jitter-dominated. Using these observations, we created a frequency, time, and polarisation-averaged pulse template. The observation archives were then fully frequency and polarisation averaged into 8\,s sub-integrations. Jitter noise is known to show a frequency-dependent evolution for bright pulsars such as PSR~J0437$-$4715 \cite[][]{2024MNRAS.528.3658K}, but for most cases, the usage of a frequency-averaged template does not affect the jitter noise \citep{2021MNRAS.502..407P}.

For our other assessments of the MPTA sensitivity (including the DM noise modelling presented in Section \ref{sec:dm} and observing strategy study presented in Section \ref{sec:uhf}), we base our analysis on the MPTA 4.5-year data set. Unlike the jitter analysis, the dataset uses sub-banded observations. A detailed description of the dataset is presented in \citet{2025MNRAS.536.1467M}. We summarise the differences below. In the 4.5-year data release, the observations are averaged into 32 frequency channels. Here, we have used 16-channel frequency resolution. To account for pulse profile evolution with frequency, 16-channel frequency-dependent pulse profile models (referred to as pulse portraits) were used as templates and cross-matched with the sub-banded observations to create the ToAs. We restrict our analysis to ToAs derived from pulse profiles with an S/N greater than $10$.
These differences are not expected to impact our assessment of sensitivity or optimisation strategies. 

\begin{table*}
\centering

\caption{Summary of dataset used for different studies in this work. The length of the data span, relevant section and assumed value of GWB amplitude for each study are shown. }
\label{tab:dataset}
\begin{tabular}{rrrrrr}
Section & Description & Data set & Data span & GWB amplitudes\\
\hline

4.1 & DM Modelling & MPTA  & 4.5 year & $1 \times 10^{-16}$\\
4.4.1 & Observing Band & Simulated & 4.5 year & $1 \times 10^{-16}$, $2 \times 10^{-15}$, $5 \times 10^{-15}$\\
4.4.2 & Single Source & Simulated & 8 year & $1 \times 10^{-16}$, $2 \times 10^{-15}$, $5 \times 10^{-15}$\\
4.5 & Scattering Noise & Simulated & 4.5 year & $ 1 \times 10^{-16}$ \\
5 & PTA Comparison & MPTA & 4.5 year & $2 \times 10^{-15}$\\

\hline
\end{tabular}

\end{table*}

\subsection{Sensitivity Curves of PTAs}

To analyse the performance of GW detectors, it is often common to produce sensitivity curves, which show the sensitivity of a detector as a function of GW frequency.  
We use the formalism presented in \citet{2019PhRvD.100j4028H} and the software package \code{hasasia} to calculate the sensitivity of a PTA to a GW signal.
We calculated the sensitivity curves in two cases: a) Continuous GWs from a non-evolving circular binary and b) an SGWB. For the continuous GW signal, a matched-filtering method can be used to determine a filter function that maximises the S/N. Since the source position and polarisation are unknown, the expected S/N averaged over the sky and inclination angle is
\begin{equation}
    \langle \rho^{2} \rangle = 2 T_{\rm obs} \int_{0}^{f_{\rm Nyq}}  \frac{S_{\rm det, \rm h}(f)}{S_{\rm det, \rm eff}(f)} df,
\end{equation}
where $S_{\rm det, \rm eff}$ is the effective strain power noise spectral density of the array, which is the weighted sum of the strain power noise spectral density of individual pulsars. $S_{\rm det, \rm h}$ is the strain power spectral density of a deterministic monochromatic source of GWs, and  $T_{\rm obs}$ is the time span of the PTA dataset \cite[][]{2019PhRvD.100j4028H}. 
The individual pulsar strain power spectral densities are estimated considering individual pulsar noise characteristics and the
response of a pulsar to a plane GW. 

For an SGWB, an optimal cross-correlation statistic is used to determine the expected S/N while considering inter-pulsar correlations. The expected S/N ratio of the optimal statistic can be expressed as
\begin{equation}
\label{eq:snr_gwb}
    \langle \rho_{\rm OS}^{2} \rangle = 2 T_{\rm obs} \int_{0}^{f_{\rm Nyq}}  \frac{S_{\rm stoch, \rm h}^{2}(f)}{S_{\rm stoch, \rm eff}^{2}(f)}  df,
\end{equation}
where $S_{\rm stoch, \rm eff}(f)$ is the effective strain power noise spectral density of the entire PTA, which is the sum of strain power noise spectral densities of pulsar pairs, including the HD correlations and $S_{\rm stoch, \rm h}(f)$ is the strain power spectral density of an SGWB \cite[][]{2019PhRvD.100j4028H}. $S_{\rm det/stoch, eff}$ can be interpreted as the detection sensitivity curve and is usually represented in terms of a dimensionless `characteristic strain', $h_{\rm c}$ such that $h_{\rm c}(f) \equiv \sqrt{f S_{\rm det/stoch,  eff}(f)}$. 

For a PTA, the pulsar noise models include white noise (typically defined as modifications to formal ToA uncertainties) and red noise processes. Both are identified through detailed modelling of the pulsar timing data sets. The timing model fit removes power at the frequencies corresponding to yr$^{-1}$ and 2\,yr$^{-1}$ because of fitting for pulsar sky position and parallax, and at lower frequencies because of fitting for a pulsar spin frequency and frequency derivative (see Figure \ref{fig:mpta_dmx}).
Similarly, power at frequencies corresponding to the reciprocal of the orbital periods is absorbed for pulsars in binary systems.
A GWB will induce a red noise signal and contribute to the spectra of all pulsars (this is GW self-noise and could manifest as CURN). The GWB amplitude that can be detected for a given configuration of the PTA at a specified S/N can be estimated using equation \ref{eq:snr_gwb}.

Here, we analyse the detection sensitivity curves for an SGWB and a single circular binary source.  
For GWB self-noise, we consider three scenarios: a pessimistic amplitude of $1 \times 10^{-16}$ \citep{2013MNRAS.433L...1S}, an intermediate value of $2 \times 10^{-15}$ (consistent with the amplitude of the signal inferred from the PPTA analysis \citep{2023ApJ...951L...6R}) and a larger value of  $5 \times 10^{-15}$  \cite[more consistent with that inferred from the MPTA,][]{2025MNRAS.536.1489M}. We summarise different datasets used in this work and the assumed GWB self-noise amplitude in each case in Table \ref{tab:dataset}.

To account for the complex noise models in MPTA, we have added additional noise processes to \code{hasasia}, in particular, DM variations and chromatic noise following the prescription given in \cite{2024ApJ...966..105A}. We include DM variations as a red noise signal following a power spectral density expressed as
\begin{equation} \label{eq:dm}
    P_{\rm DM}(f) = \frac{A_{\rm DM}^{2}}{12 \pi^{2}} \bigg(\frac{f}{f_{c}}\bigg)^{-\gamma_{\rm DM}} \bigg(\frac{\nu}{\nu_{\rm ref}}\bigg)^{-4}
\end{equation}
where $A_{\rm DM}$ and $\gamma_{\rm DM}$ are the amplitude and spectral index of DM noise power law spectrum, respectively, at a reference frequency $f_{c}$ at $1\rm yr^{-1}$.
Similarly, for chromatic noise addition, we assume the power spectral density of the form
\begin{equation} \label{eq:cn}
    P_{\rm chrom}(f) = \frac{A_{\rm chrom}^{2}}{12 \pi^{2}} \bigg(\frac{f}{f_{c}}\bigg)^{-\gamma_{\rm chrom}} \bigg(\frac{\nu}{\nu_{\rm ref}}\bigg)^{-2\beta}
\end{equation}
where $\beta$ is the chromatic index, a free parameter encompassing the frequency-dependent multi-path propagation delays.
We did not incorporate the impact of stochastic solar wind electron density variations in our analysis.

\section{Jitter noise} \label{sec:3}
\subsection{Methodology}

To measure jitter noise for each pulsar in our sample, we first generated a frequency, time, and polarisation-averaged template, and then calculated ToAs for frequency-averaged  8\,s sub-integrations. 
These ToAs are then fitted using a previously derived timing model \citet{2025MNRAS.536.1467M} with all parameters fixed except spin frequency to estimate the weighted-rms ($W_{\rm rms}$) of the timing residuals. We only fitted for spin frequency to absorb red noise-like signals so that the remaining scatter can be attributed to jitter noise. As we have used frequency-averaged observations, we did not fit for DM.

On short timescales, the error in ToA can be described as
\begin{equation}
    \sigma_{\rm tot}^{2} = \sigma_{\rm S/N}^{2} + \sigma_{\rm J}^2 + \sigma_{\rm DISS}^{2}
    \label{Eq:toterr}
\end{equation}
where $\sigma_{\rm S/N}$, $\sigma_{\rm J}$ and $\sigma_{\rm DISS}$ are the errors associated with the radiometer noise, jitter noise, and diffractive scintillation noise. The latter term arises from the stochasticity in the pulse broadening due to the turbulent IISM and is expected to contribute most to highly scattered pulsars that usually possess high DMs \citep{2010arXiv1010.3785C}. Here we have neglected this because previous studies have suggested that for most pulsars the effect of $\sigma_{\rm DISS}$ is found to be smaller than the $\sigma_{\rm J}$ and $\sigma_{\rm S/N}$ in a comparable sample \citep{2024ApJ...964....6W}.

To measure jitter noise, we first simulated idealised ToAs with ToA uncertainties the same as the observations using \code{TEMPO2} following the practice outlined in \cite{2021MNRAS.502..407P}, creating $1000$ realisations of the simulated dataset using the high flux density observation.
We use simulations as they allow us to set robust limits and confidence intervals,  which is especially important as we are setting jitter limits from short observations consisting of a few independent arrival times.\footnote{For example, for many of the pulsars, the jitter measurements would be derived from observations with $256$\,s duration and $8$\,s sub-integrations.}
Therefore, uncertainty due to jitter noise in a sub-integration length $T_{\rm sub}$ can be estimated to be   
\begin{equation}
   \sigma_{\mathrm{J}}^{2} (T_{\mathrm{sub}}) = \sigma_{\mathrm{obs}}^{2} (T_{\mathrm{sub}}) - \sigma_{\mathrm{sim}}^2 (T_{\mathrm{sub}})
    \label{Eq:jitter}
\end{equation} 

Similar to \cite{2021MNRAS.502..407P}, we report the detection of jitter noise if the variance in the observations is greater than that measured in 95\% of the realisations of idealised data sets. 

Jitter measurements are often reported as those that would result from one-hour duration integrations. Since jitter is expected to scale with the number of pulses being averaged as $1/\sqrt{N_p}$, we can estimate jitter in an hour to be
\begin{equation}
\label{eq:sing_jitter}
    \sigma_{\mathrm{J}} (\mathrm{hr}) = \frac{\sigma_{\mathrm{J}} (T_{\mathrm{sub}}) }{\sqrt{3600/T_{\mathrm{sub}}}}.
\end{equation}

\subsection{Results}
We have measured jitter noise in $41$ MSPs (20 of which are new) and report upper limits for the remaining 48 MSPs as presented in Table \ref{tab:jitter}.
For the brightest MSP, PSR~J0437$-$4715, we measured jitter noise of 52(7)\,ns in 1\,hr. The lowest jitter level is shown by PSR~J2241$-$5236 at 3.5(3)\,ns in 1\,hr, consistent with the value obtained by \cite{2021MNRAS.502..407P}. Our MSP sample overlaps with previous MSP jitter studies such as \cite[e.g.,][]{ 2014MNRAS.443.1463S, 2019ApJ...872..193L, 2021MNRAS.502..407P, 2024ApJ...964....6W}.  A comparison of results is presented in Table \ref{tab:jitter}. Our jitter measurements are consistent with the previous studies at an $\sim 1\sigma$ level. 
 
\begin{figure}
\centering
\includegraphics[scale=0.6]{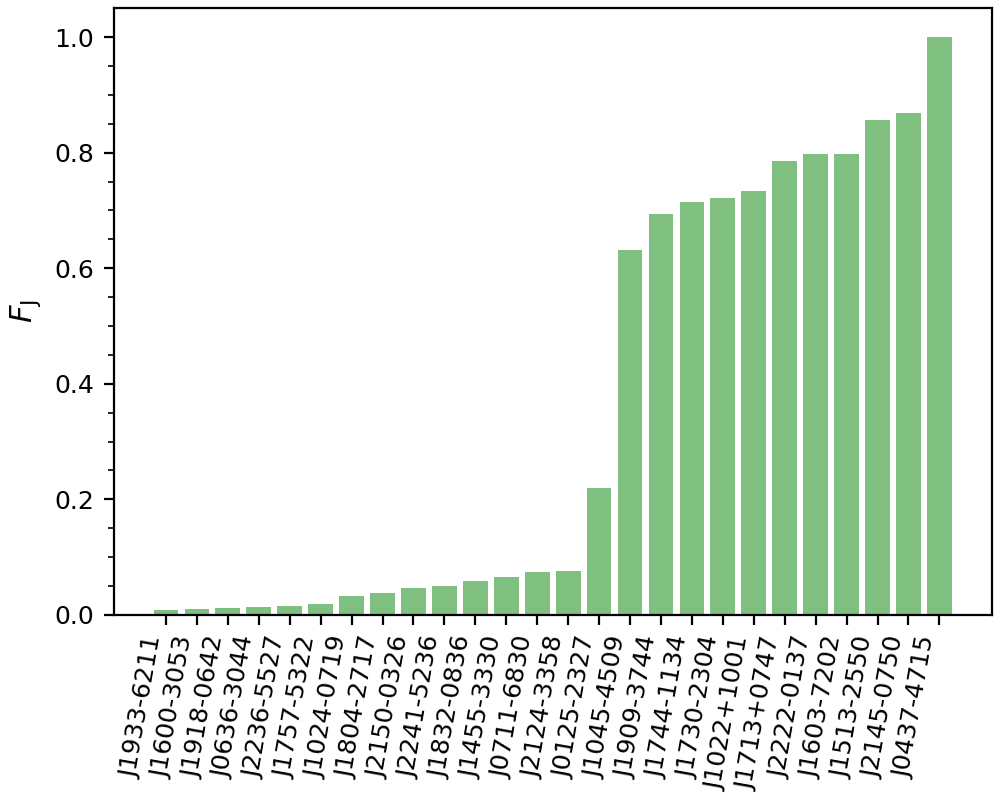}
\caption{Fraction of jitter-limited observations ($F_{\rm J}$) per pulsar in MPTA. Out of the 41 MSPs, 15 pulsars are not included in this analysis as none of the observations were jitter-limited according to our definition. The pulsars are sorted by the fraction, and PSR~J0437$-$4715 has 100\% observations that are jitter-limited. }
\label{fig:frac_obs}
\end{figure}

Propagation of pulsar radio emission through the IISM causes flux density variations, i.e. interstellar scintillation. As such, it is likely that jitter noise is the only limiting noise source for a fraction of our observations ($F_{\rm j}$), where the flux density is enhanced.
We assume that an observation is jitter-limited if the ToA uncertainty ($\sigma_{\rm S/N}$), measured from a fully time and frequency-averaged observation, is less than the jitter noise in the epoch-wise observing length. In Figure \ref{fig:frac_obs}, we plot $F_{j}$ for pulsars in which we have detected jitter with MeerKAT.  We find that, for the nearest and brightest MSP, PSR~J0437$-$4715 \citep{2024MNRAS.530..287S}, all of the observations are jitter-limited,  whereas for 15 pulsars (36\%), none of the observations are jitter-limited. 
\begin{figure}
\centering
\includegraphics[width=\columnwidth]{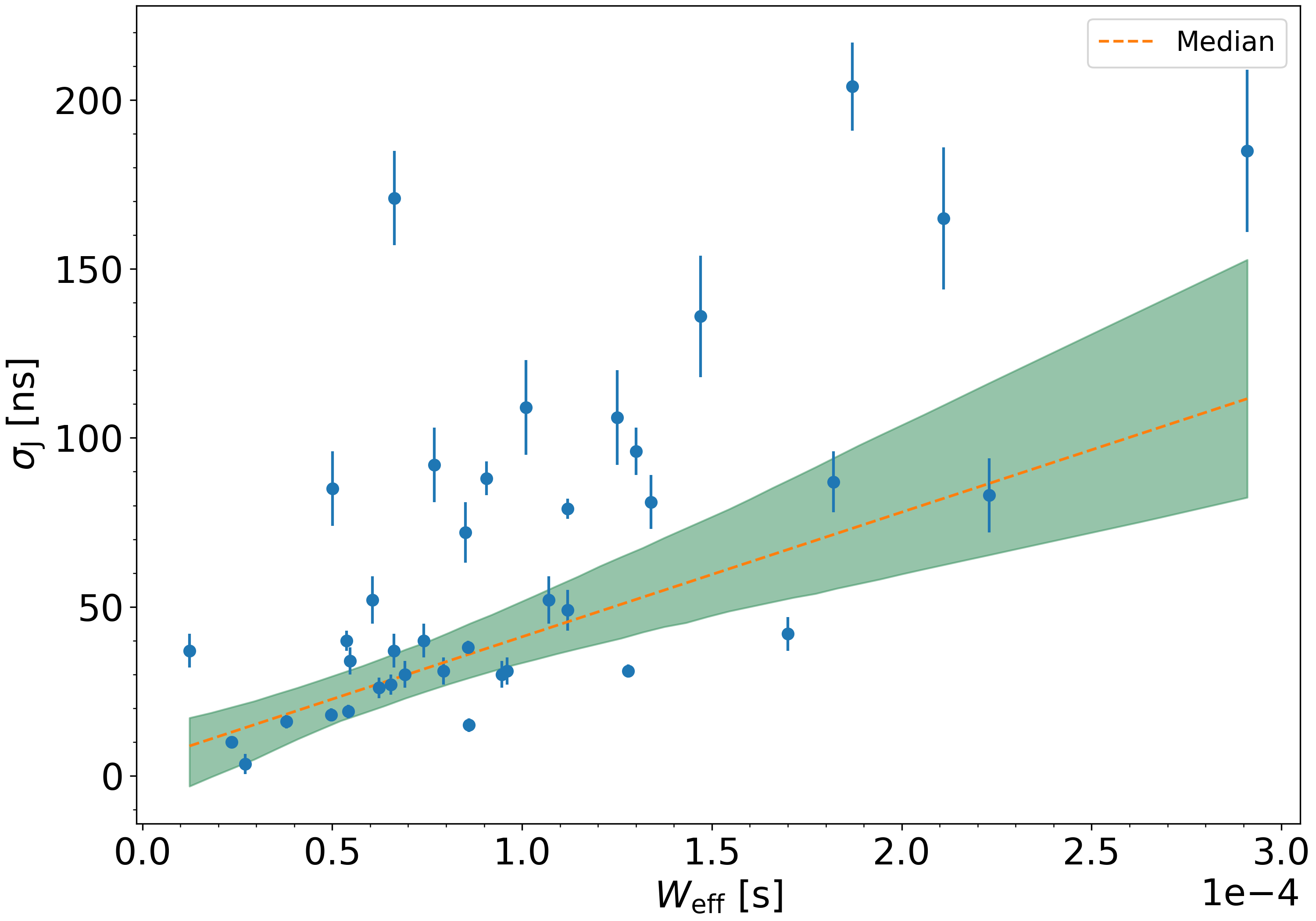}
\caption{Comparison of  $W_{\rm eff}$ and $\sigma_{\rm J} (\rm hr)$ for the MPTA pulsars. The median relationship is $36.9(8.1) \times (W_{\rm eff}/10^{-4} ) + 4.2(2.7)$ and the 2$\sigma$ uncertainty regions for the likelihood fit are plotted. 
}
\label{fig:weff_jit}
\end{figure}

We have also searched for a correlation between pulse profile effective width, which is a proxy for the sharpness of the pulse profile $W_{\rm eff}$ and jitter noise.
We calculate $W_{\rm eff}$ using pulse portraits, generated similar to those presented in  \cite{2023MNRAS.519.3976M} using the \code{PulsePortraiture} method \cite[][]{2019ApJ...871...34P}. 
We calculate $W_{\rm eff}$, using the relationship given in \cite{2010arXiv1010.3785C}: 
\begin{equation} \label{Eq:weff}
    W_{\rm eff} = \frac{\Delta t}{\sum_{\rm j} (U_{\rm j+1} - U_{\rm j})^{2}}
\end{equation}
where $U_{\rm j}$ is the profile (normalised to peak flux density) and $\Delta t $ is the time resolution of the pulse profile.  
We modelled $\sigma_{\rm J}$ for the MSPs using a linear relationship between $W_{\rm eff}$ and $\sigma_{\rm J}$ as shown in Figure \ref{fig:weff_jit}. We derive a median and 2$\sigma$ region dependence for a linear fit model using \code{bilby} \citep{bilby_paper}. In our model, we scaled the uncertainties by adding two parameters; one multiplied by the errors and the other added in quadrature in order to make the likelihood better describe the data, but we plot the data using the original uncertainties.
We also considered a power law relationship between $W_{\rm eff}$ and $\sigma_{\rm J}$ (i.e.,  $\sigma_{\rm J} \propto W_{\rm eff}^\alpha)$. We found $\alpha=0.95\pm 0.09 $, which is consistent with a linear relationship.  

Using our larger sample of pulsar jitter measurements, we searched for correlations between jitter noise and other pulsar intrinsic and extrinsic parameters, including pulse width ($W_{50}$), sharpness of pulse profile ($W_{\rm eff}$), pulsar period ($P$), and flux density at 1400\,MHz ($S_{1400}$) of the pulsar. We use values of $W_{50}$ reported in the ATNF pulsar catalogue \citep{2005AJ....129.1993M}. 
To be consistent with previous studies, we also converted $\sigma_J(\rm hr)$ to single pulse jitter $\sigma_{\rm J, 1}$ ,( replacing $T_{\rm sub}$ as $P$ in equation \ref{eq:sing_jitter} ) and the jitter parameter $k_{\rm J} = \sigma_{\rm J, 1}/P$, which removes bias related to pulse period. 
We use the Spearman correlation coefficient (R) as our test statistic and use its p-values to test for statistical significance. We summarise our findings and compare them with previous works in Table \ref{tab:jit_corr}. 
Our results suggest that there is significant correlation between pulse jitter and the $W_{\rm eff}$ of a pulsar.
However, when normalising by pulse period, the correlation becomes insignificant. 

\begin{table}
\centering

\caption{Spearman correlation coefficient between jitter noise and pulsar parameters. The first four columns denote jitter noise, pulsar parameters, correlation coefficient and associated p-value. The last column shows correlation coefficients reported in previous works and associated p-value if available. Here L+19, P+21 and W+24 refers to \citet{2019ApJ...872..193L}, \citet{2021MNRAS.502..407P} and \citet{2024ApJ...964....6W} respectively.  We note that both the $W_{\rm eff}$ and R relation used by \cite{2019ApJ...872..193L} differed from other works.}
\label{tab:jit_corr}
\begin{tabular}{rrrrrr}
\hline

Jitter  & Parameter& R & p-value & Previous work\\
\hline
$\sigma_{\rm J} (\rm hr)$ & $W_{50}$ & 0.46 & 0.01& 0.64(P+21)\\
$\sigma_{\rm J,1}/P$ & $W_{50}/P$& 0.18& 0.30& 0.62(L+19), 0.82(W+24)\\
$\sigma_{\rm J} ( \rm hr)$ & $W_{\rm eff}$ & 0.65 & $8\times 10^{-6}$& 0.64 (P+21) \\
$\sigma_{\rm J,1}/P$ & $W_{\rm eff}$ & -0.18& 0.28& (L+19) 0.2(0.45)\\
$\sigma_{\rm J,1}$ & $W_{\rm eff}$ & 0.49& $1.6 \times 10^{-3}$& \\
$\sigma_{\rm J} ( \rm hr)$ & $S_{1400}$& -0.05& 0.79&\\

\hline
\end{tabular}
\end{table}

\begin{figure}
\centering
\includegraphics[width=\columnwidth]{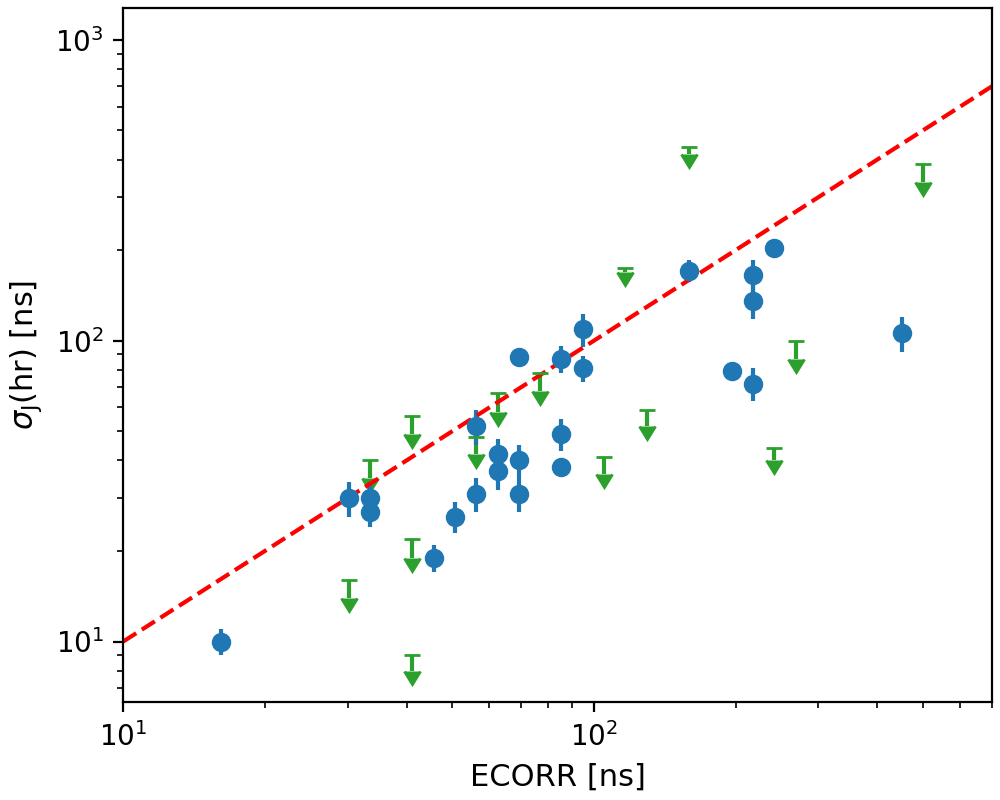}
\caption{Comparison of ECORR parameters derived in MPTA noise analysis and jitter values estimated in this work. The blue points and green points denote direct measurements of jitter noise and upper limits, respectively. Jitter measurements are systematically lower than the ECORR values. The red dashed line corresponds to a 1:1 relationship between $\sigma_{\rm J} (\rm hr)$ and ECORR.}
\label{fig:ecorr_jit}
\end{figure}

We also investigated the connection between jitter noise derived from our single-epoch observations and that derived from long-term timing observations.
In the long-term pulsar noise analysis, ECORR describes the uncertainties that are correlated between contemporaneous arrival times for each observing backend \citep{2015ApJ...813...65N}. This term includes pulse jitter noise but could also include other noise terms correlated within an observation but uncorrelated between epochs, such as broadband RFI or IISM effects \citep{2010ApJ...717.1206C}. In Figure \ref{fig:ecorr_jit}, we compare the ECORR values estimated by \citet{2025MNRAS.536.1467M} to the jitter noise values estimated in this work. 
While for some pulsars, there is the expected linear correlation between jitter noise and ECORR, for the majority, it is clear that ECORR overestimates jitter noise. Previously, \cite{2016ApJ...819..155L} reported a similar overestimation of jitter noise in the NANOGrav 9-yr dataset. If the source of excess noise measured in ECORR can be mitigated, then using the measured jitter noise values instead of ECORR will reduce the number of free parameters and could improve the sensitivity of PTAs. To demonstrate this, we compare the sensitivity of the MPTA assuming the ECORR values derived from long-term timing against the sensitivity assuming ECORR corresponding to  $\sigma_{\rm J}$. We see an improvement of 8\% when using jitter noise as the ECORR value for a GWB amplitude of $1 \times 10^{-16}$, the scenario that would show the greatest improvement.
For the rest of the work, we will use ECORR derived from long-term timing in our array sensitivity calculations.

\begin{figure*}
\centering
\includegraphics[width=\textwidth]{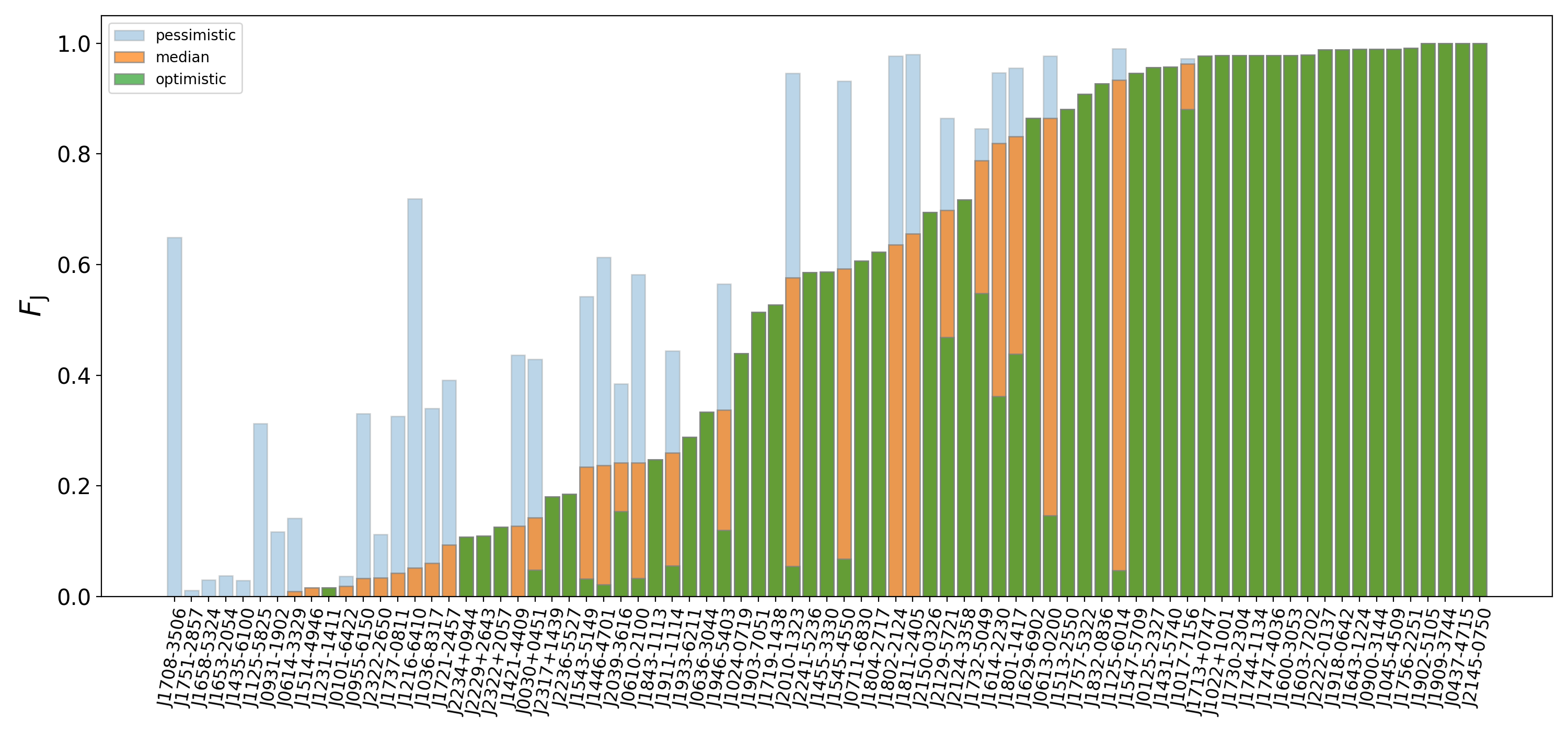}
\caption{Fraction of jitter-limited observations ($F_{\rm J}$) for each pulsar with a telescope with SKA-Mid sensitivity. Jitter noise is measured in this work for 41 MSPs. For the MSPs without jitter measurements, we have used the relation between $W_{\rm eff}$ and $\sigma_{\rm J}(\rm hr)$ derived from  Figure \ref{fig:weff_jit} to estimate the jitter. Hence, three fractions are shown: one using the median relation, and the other two showing the 2$\sigma$ region to define optimistic (green) and pessimistic (blue) scenarios for the jitter noise in these pulsars. The fractional values are sorted by the median method. Many high-precision MSPs in SKA-Mid will be completely jitter-limited if observed with the entire array.}
\label{fig:ska}
\end{figure*}

Jitter noise will have a larger contribution for MSPs observed with even more sensitive telescopes, such as the SKA-Mid telescope. The design sensitivity for the SKA-Mid (AA4, comprising 197 antennas) is forecast to be a factor of $4.1$ more sensitive than the MeerKAT telescope \citep{2019arXiv191212699B}, and it will have an L-band observing system with comparable bandwidth to MeerKAT. We simulated SKA-Mid observations for our MSP sample by scaling the ToA uncertainties by their sensitivities and repeated the above procedure to calculate $F_{j}$. For the MSPs without a $\sigma_{\rm J}$ measurement, we make a prediction based on the correlation between $W_{\rm eff}$ and $\sigma_J$ described above.  
Based on these relations, we simulated the SKA-Mid $\sigma_{\rm J}$ using the median values for each MSP, and pessimistic (jitter noise $+ 2\sigma$ above the median) and optimistic ($2\sigma$ below the median) values. The fraction of observations that would be jitter-limited with the SKA-Mid is displayed in Figure \ref{fig:ska}, with pulsars sorted by the `median' model. In the pessimistic case, we can see spikes in $F_{\rm j}$ for certain pulsars. We note that for these pulsars, ToA uncertainties are more Gaussian distributed than the other pulsars. Therefore, when a certain scenario is assumed, e.g. pessimistic,  a higher number of observations becomes jitter-limited compared to the remaining pulsars. Owing to the increase in sensitivity, more pulsars are completely jitter-limited with SKA-Mid, including one of the most precisely timed MSPs, PSR~J1909$-$3744.

Finally, we considered whether it is possible to improve array sensitivity by increasing the integration time for the jitter-limited pulsars and reducing the time of other pulsars such that the total observing hours remain the same. We created a jitter-based sub-array by choosing the 11 most highly jitter-limited pulsars with the MPTA, as shown in Figure \ref{fig:frac_obs}. In particular, we considered a scenario where these pulsars were observed 10 hrs in total per observing session using 16 dishes of MeerKAT at L-band, with the remaining pulsars observed using 48 dishes. Considering a typical observation length of 256\,s, this sub-array method will increase the ToA uncertainty in a jitter-limited pulsar by a factor of 1.11 while decreasing the jitter noise by a factor of 3.6. For such a jitter-based sub-array, we did not find an improvement in sensitivity relative to the standard strategy. If we compare the change in sensitivity of an array just made from the 11 jitter-limited pulsars, then observing longer increases sensitivity by 20\%, but is counterbalanced by the loss in sensitivity (15\%) for the remaining pulsars due to the reduction in the number of dishes and therefore telescope gain. 
\section{Optimisation strategies} \label{sec:4}

\subsection{DM noise modelling}\label{sec:dm}
 
As the pulsar radiation traverses different lines of sight due to the motion of the pulsar, the Earth, and the IISM, it encounters different electron column densities. As a result, the DM varies with time, introducing time and frequency-correlated delays proportional to $\nu^{-2}$, where $\nu$ is the radio frequency at which ToAs are calculated. There are two common ways to mitigate and fit for the effects of DM variations. Many PTA data analyses model the DM variations as a stochastic correlated GP with a red noise power spectral density given by equation \ref{eq:dm}. The DM GP power law amplitude and spectral index are inferred from the timing residuals with a series of Fourier basis functions.
An alternative approach, in particular used by the NANOGrav collaboration, is to use a piece-wise constant function to model DM variations \citep{2013ApJ...762...94D}. Owing to their multi-epoch observation with different frequency band receivers, they assume that DM does not change over a period of $\sim$14 days (although this number is tuned for each pulsar) and fit for the DM variations relative to a fiducial value over their observation time span.  This method is commonly referred to as the DMX method. 
Here, we analyse our dataset based on both these DM variation models to assess which model provides higher sensitivity to GWs. 

To compare the DM GP and DMX methods, we use the 4.5-year MPTA data set as described above. 
To account for DMX, we needed to choose a DM sampling cadence, which is included in the pulsar timing ephemerides.
We choose DM epochs centred around each observation of a pulsar. This is appropriate as all MPTA observations are taken in the same band, so all observations have the same frequency coverage. Using the standard timing software \code{TEMPO2}, a unique DM offset is fitted for all the ToAs in each epoch-based window using the $\nu^{-2}$ dependence while the global DM is fixed.

\begin{figure}
\centering
\includegraphics[width=\columnwidth]{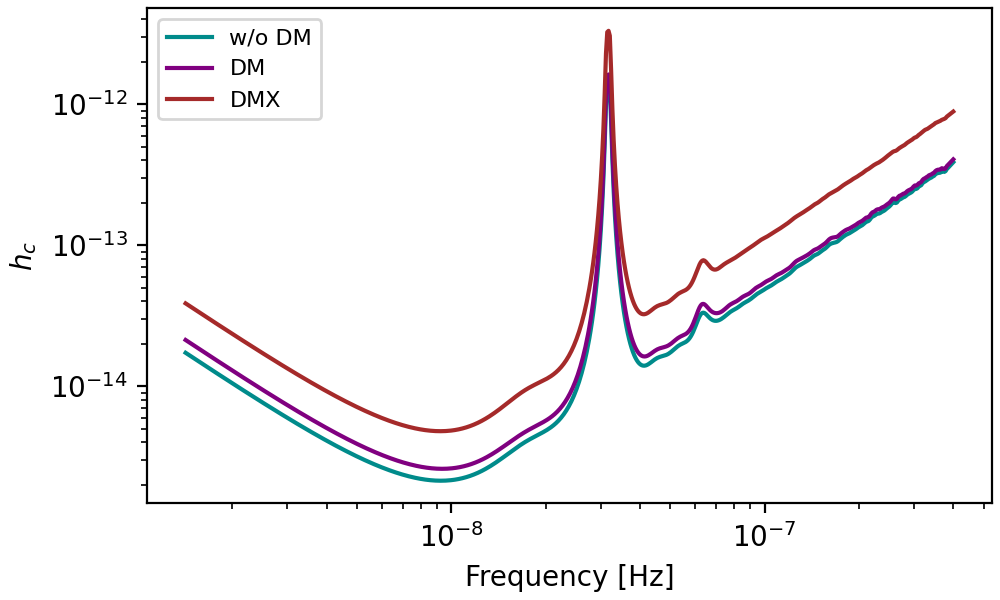}
\caption{Comparison of sensitivity curves for the MPTA with alternate DM noise modelling methods. The cyan curve corresponds to white noise, red noise and chromatic noise, the purple curve additionally includes DM as a GP, whereas the crimson curve includes DMX. The DMX implementation reduces the sensitivity by 44\% compared with the DM GP model. }
\label{fig:mpta_dmx}
\end{figure}

For all of the MSPs in the MPTA, we studied the effect of the choice of DM noise modelling methodology on the sensitivity of the array for a range of DM GP properties. For individual MSPs, we varied the DM noise amplitude (in the range $-20 < \log_{10} A_{\rm DM} < -10$) and spectral index around the value ($\gamma_{\rm DM} \pm 2$) inferred from the 4.5-year noise analysis \citep{2025MNRAS.536.1467M}, and estimated the minimum SGWB amplitude that can be detected for each combination. The motivation behind this was to determine if pulsars with lower dispersion measure were more sensitive to the choice of DM modelling approach than the higher dispersion measure pulsars (or vice versa). We found that the GP method is more sensitive to an SGWB in all cases. We then compared the entire array sensitivity with respect to the choice of DM noise model.  The noise sensitivity curve for both cases is shown in Figure \ref{fig:mpta_dmx}. We found that the DM GP method improves the array sensitivity by 44\% compared to the DMX method. 
We attribute this decrease in sensitivity to the large number of additional parameters added to the timing model when DMX is used.  This was also observed for an analysis of a similar DM correction scheme described in \cite{2013MNRAS.429.2161K}. However, we note that there are many caveats to choosing one DM model over the other. In the DM GP model, DM variations as a function of time, DM(t), are modelled as a finite sum of sine and cosine functions, which are truncated to lower computational costs. Therefore, the recovered amplitude and spectral index are approximations of the true solutions \citep{2014PhRvD..90j4012V}. Apart from that, the DM GP model assumes stationarity of the signal, whereas extreme scattering events can make this assumption invalid \citep{2015ApJ...808..113C, 2018MNRAS.474.4637K}. The DMX model can potentially better account for this non-stationarity, whereas the DM GP model better models the long-term correlated noise due to DM.

\subsection{Observing strategies with the multi-frequency receivers of MeerKAT} \label{sec:uhf}
The flexibility of MeerKAT motivates optimising sensitivity through the choice of receiver or sub-array.
In order to assess the sensitivity, it is first necessary to estimate the achievable timing precision in other observing bands.
Due to the lack of timing campaigns with the UHF and S-band receivers, it was necessary to base timing precisions on simulations extrapolated from the L-band observations.
Using the 16 frequency channel observations, we first created idealised ToAs with added Gaussian noise using \code{TEMPO2}. This way, the ToA errors were consistent with the original observation. In our simulations, we assumed the entire bandwidths at the UHF band and S band are free from RFI. The 544\,MHz and $\sim$ 870\,MHz of available bandwidth of UHF and S-band systems were divided into 10 and 16 channels, respectively, to match the channel bandwidth of the L-band system. 
To simulate the ToA error ($\sigma_{\rm ToA}$) for the different frequency receivers, we started with an assumed relationship between timing precision and effective width ($W_{\rm eff}$): 
\begin{equation} 
    \sigma_{\rm ToA} \propto \frac{W_{\rm eff}}{\rm S/N} (T_{\rm sys} + T_{\rm sky}),
    \label{Eq:snr}
\end{equation}
where $T_{\rm sys}$ and $T_{\rm sky}$ are the system and sky temperature, respectively.
The $W_{\rm eff}$ , S/N of observations, $T_{\rm sys}$ and $T_{\rm sky}$ change with frequency, so this relation can be used to extrapolate the ToA uncertainties at different frequencies. Here, we have considered $T_{\rm sys}$ to be fixed at 18\,K for L-band and 23\,K for UHF band, which correspond to 1400\,MHz and 815\,MHz (centre frequency of UHF-band), respectively. For each pulsar, we calculated $T_{\rm sky}$ based on the \cite{1982A&AS...47....1H} 408\,MHz all sky brightness temperature map for each pulsar position, and scaled it as $\nu^{-2.6}$ \citep{1987MNRAS.225..307L} at each frequency sub-band, where $\nu$ is the radio frequency. For the S band,  $T_{\rm sky}$ will have negligible effects on sensitivity and hence can be neglected. 

Since the flux density of a pulsar is proportional to the observed S/N, we used the known measured spectral indices of the pulsars, reported in \cite{2023MNRAS.526.3370G}, to extrapolate flux density values at different frequencies.
The pulse width of a pulsar is expected to increase because of scattering and intrinsic pulse-width variations \citep{ 1975A&A....38..169S,1987AuJPh..40..557S}.
However, \cite{2023MNRAS.519.3976M} showed that  $W_{\rm eff}$ can exhibit diverse and sometimes non-monotonic evolution with pulse frequency for MPTA pulsars. Using measurements of $W_{\rm eff}$ as a function of frequency in L band, revealed that at UHF frequencies, some pulsars can have three times higher or lower $W_{\rm eff}$ in comparison to L band.
To model the frequency dependence of $W_{\rm eff}$, we considered power-law, quadratic, and cubic polynomial relationships, and selected the model that best described the L-band observations as measured by the lowest chi-squared value. Of the 83 MSPs, 62 pulsars followed a quadratic relation between $W_{\rm eff}$ and frequency with the remaining having either power-law or cubic relationships.
 
For each epoch, the ToA uncertainties were estimated using equation \ref{Eq:snr} and the width and flux density relations with frequency as derived above.
As a result, the ToA uncertainty remains constant across all the epochs but changes across frequency. 
We note that this neglects the effect of scintillation modulating the S/N of the observations, and hence the effects of variable timing precision. 
To make the comparisons between bands fairer,  we also simulated L-band observations based on these extrapolations. Between the observed L-band dataset and the simulated dataset, the sensitivity differs by just 5\%. This suggests that incorporating scintillation into our simulated data set would only marginally change the results.

\subsection{Confirming our UHF predictions}

While we did not have sufficient observations to measure timing precision in the UHF-band, to assess if our projections were reasonable, we conducted a single epoch of observations of the MPTA pulsars in March 2024 with MeerKAT using the UHF receiving system. 
We investigated whether our predictions for $W_{\rm eff}$ and flux density matched the observations.
The tests were impacted by only having a single epoch. For low DM pulsars, diffractive scintillation introduces an additional systematic in the uncertainties of the flux density measurements and can make it difficult to assess how the pulse profile is evolving with frequency. For similar reasons, it was also difficult to make reliable measurements for the weaker pulsars in the MPTA sample.
As a result, template creation with the \code{PulsePortraiture} method did not always converge. 
Instead, to estimate $W_{\rm eff}$ for the UHF observations, we averaged the observations to 10 frequency channels and smoothed the profiles. For many low DM pulsars, we found a deviation in the projected flux density and frequency power law due to scatter broadening of the pulse profile at the lowest frequencies. In most cases, our predictions for $W_{\rm eff}$ as a function of frequency were consistent with observations. However, for 25 MSPs, the observation followed either a broken power law or a simple power law with an amplitude and index different from those derived using L-band observations. We modified our models and simulated the dataset based on these new observations. 

\subsection{Sub-arrays with MeerKAT} \label{sec:uhf_subarray}
\subsubsection{Observing strategies for SGWB detection}

\begin{figure*}
\centering
\includegraphics[width=18cm, height=5cm]{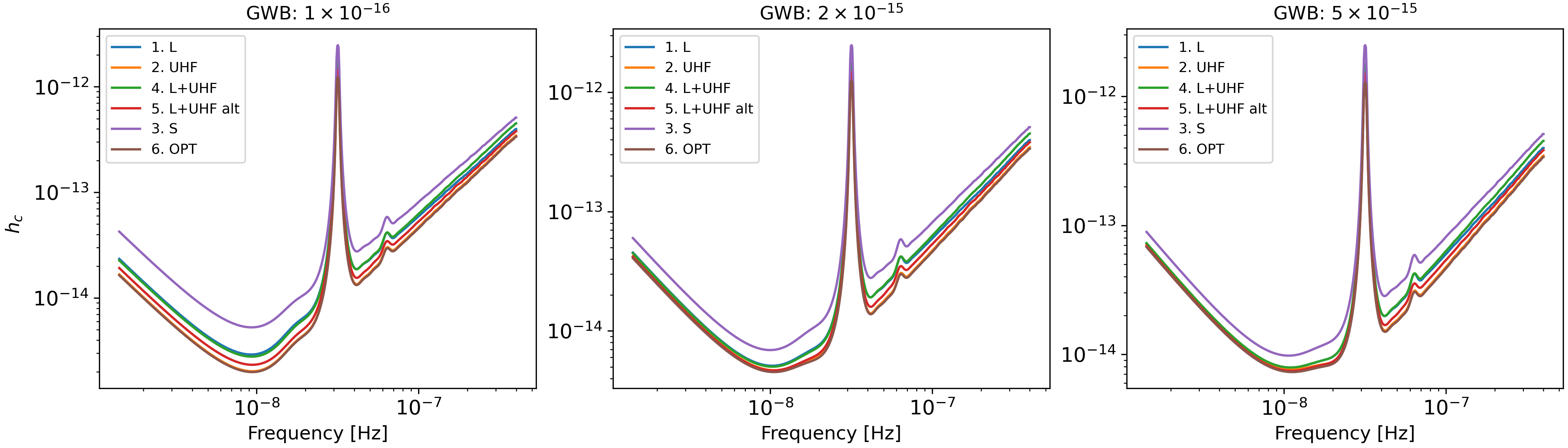}
\caption{Sensitivity curves for MPTA observing strategies. The curves show the following strategies: L-band (L), UHF-band (UHF), S-band (S), observing with L and UHF simultaneously (L+UHF), observing with L and UHF-band alternatively (L+UHF alt) and choosing the optimal method out of the above for each pulsar (OPT). The three sub-panels show the same curves for three values of SGWB self-noise: $1 \times 10^{-16}, 2 \times 10^{-15}, 5 \times 10^{-15} $.}
\label{fig:mpta_arr}
\end{figure*}

Using the L-band, UHF-band, and S-band simulated data sets, we considered the following observing strategies and compared the resulting sensitivity:
 \begin{enumerate}
\item Observe with L-band exclusively
\item Observe with UHF-band exclusively
\item Observe with S-band exclusively
\item Observe with UHF and L-band simultaneously with 32-antenna sub-arrays 
\item Observe with UHF and L-band alternatively, maintaining the same overall cadence
\item Observe with the optimal strategy out of cases 1-5 for each pulsar
 \end{enumerate}
For case (4), we combined the simulated L-band and UHF-band data sets and
multiplied the ToA uncertainties by a factor of two to account for the
loss in gain due to the reduction in the number of antennas in each sub-array. 
For case (5),  the dataset was created by discarding alternate epochs of observations from the combined L and UHF-band datasets. 
For case (6), we chose the best method
of observation for each pulsar out of the remaining observing strategies.

For each strategy, we use \code{hasasia} to generate the sensitivity curves
for individual MSPs assuming a low ($1 \times 10^{-16}$), medium ($2 \times 10^{-15}$) and high ($5 \times 10^{-15}$) amplitude for the SGWB. Each curve assumes DM GP and ECORR model for which we have used MPTA 4.5 year noise models given in \citet{2025MNRAS.536.1467M}. For the majority of
MSPs, UHF-only observations are the preferred strategy, with case (5) being
the second most preferred. S-band observations were the most sensitive for 13 MSPs. Many of these pulsars have significant scattering noise contributions, which dominate at lower frequencies and decrease the timing precision at low frequencies.
Figure \ref{fig:mpta_arr} shows the sensitivity curves for the entire array based on the above six observing strategies. 
Out of the six observing strategies, UHF exclusive observations (case 2)  provide 30\% improvement over
L-band exclusive observations. S-band exclusive observations are a factor of $1.8$ less sensitive than L-band observations for an SGWB amplitude of $1 \times 10^{-16}$. This is expected given the low flux density observations at the S-band and steep spectral indices. The optimal strategy is nearly identical to the UHF exclusive strategy. To understand if we could mitigate frequency-dependent noise processes in UHF and L band equivalently, we calculated the error in infinite frequency ToAs using the equation in \cite{2010arXiv1010.3785C}. For the simulated dataset of PSR~J0437--4715, we found the error in infinite frequency ToA in UHF-band to be 10\% better than L-band.
We conclude that trade-offs between timing precision (impacted by sensitivity and pulse width) and frequency coverage result in similar ability to correct for DM in  UHF and L-band. 

Among the mixed L-band and UHF-band
observations, alternate observations show higher sensitivity than 
simultaneous sub-array observations using half of the dishes in each array due to favourable linear ToA uncertainty scaling with gain in contrast to the square root improvement that comes from increasing observing time.
This suggests that the increased ToA precision is more important than simultaneous observations, assuming propagation effects can be optimally modelled, an assumption we investigate further below. We also note that at higher GW frequencies, the slope of the sensitivity curve for case (4) is higher than for case (5) as the use of fewer antennas increases the white noise contribution. 
However, for practical purposes, simultaneous observations might be beneficial in characterising low-frequency effects like chromatic noise, especially if they vary on the timescale of the observing cadence.

As we increase the SGWB self-noise, the lower frequencies become GW-dominated, and the amount of improvement decreases between the UHF-only and L-band only observations. For SGWB amplitude of $2 \times 10^{-15}$ and $5 \times 10^{-15}$, UHF is only 8\%  and 2\% more sensitive  than  L-band observation, respectively.  We also note these improvements assume perfect correction of chromatic effects, an assumption we test below.


\subsubsection{Observing strategies for single source detection}

Continuous GWs from non-evolving single SMBHBs could dominate over a stochastic background, particularly at higher frequencies ($f \gtrsim$ a few 10s of nHz) \citep{2008MNRAS.390..192S, 2009MNRAS.394.2255S}.
We investigated whether alternate observing strategies in further observations would result in improved sensitivity by estimating the sky-averaged sensitivity of the array with five years of observations with the current MPTA observing cadence (as the MPTA has had a consistent observing strategy for five years up to the end of 2024) and then adjusting the observing strategy for the subsequent three years. We chose to modify the cadence for three years, as that marks the beginning of science verification for 64 SKA-mid dishes, which at the time of writing this paper, was expected to be combined with the 64 MeerKAT antennas in late 2028.

\begin{figure}
\centering
\includegraphics[width=\columnwidth]{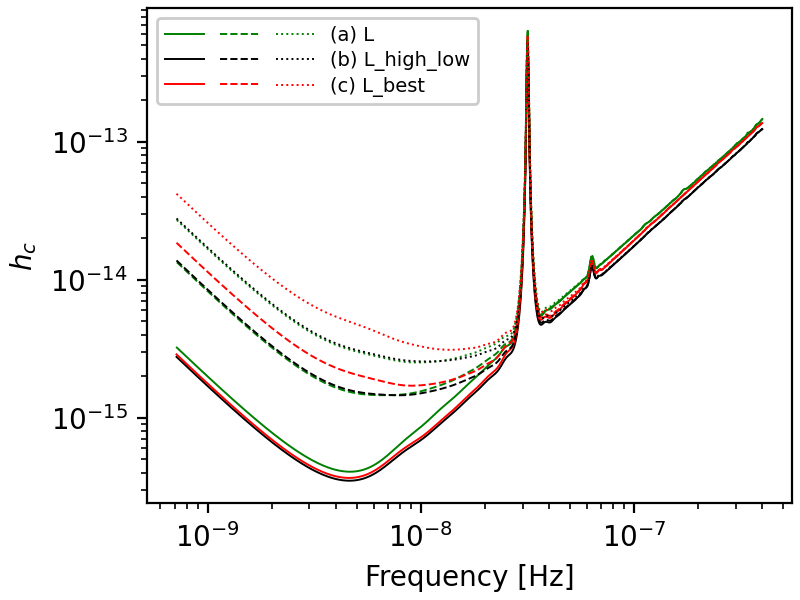}
\caption{ Sky-averaged sensitivity of MPTA for a single continuous GW source.  We consider cases where the SGWB self-noise amplitude is $1 \times 10^{-16}$ (solid), $2 \times 10^{-15}$ (dashed), and $5 \times 10^{-15}$ (dotted). We compare the sensitivities for different observing strategies: (a) L-band simulated 8 year dataset (green) (b)  L-band 5-year dataset + 3-year dataset with high-cadence observations of the 20 best pulsars, and low-cadence observations of the remaining pulsars (black) (c) L-band 5 year dataset of 20 best pulsars + 3 year dataset of high-cadence observations (red). Case (a) and (b) provide similar sensitivities for SGWB and single sources in the case of higher GW self-noise, whereas case (b) provides best sensitivity for single source detection without reducing sensitivity towards an SGWB.   }
\label{fig:mpta_sc}
\end{figure}

The strategies we considered with simulated datasets were,
Case (a) 5 year MPTA observations + 3 years observations with the current MPTA strategy \\
Case (b) 5 year MPTA observations + 3 year observations of 20 best MSPs with a factor of four increased cadence while reducing the observation cadence of remaining MSPs, keeping the telescope time fixed \\
Case (c) 5 year MPTA observations + 3 year observations of 20 best pulsars with a quadrupled cadence \\
Case (d) 5 year MPTA observations + 3 year observations of 5 added MSPs with 4\,us ToA error without any red noise. \\ 
In Figure \ref{fig:mpta_sc}, we plotted the sensitivity curves for the  single source strategies.
The high-cadence observations of the best pulsars i.e. case (b) provide the highest sensitivity at the higher frequencies, with a $\sim$14\% improvement over case (a). Timing the best MSPs (case c) provides similar sensitivity towards single source detection as timing the entire array. The addition of five pulsars (case d) marginally improves the sensitivity compared to case (a), therefore, we did not include the corresponding curve in the Figure \ref{fig:mpta_sc}. We speculate these pulsars are not strongly contributing to the array's sensitivity because of their higher ToA errors and shorter observing spans. For a more realistic scenario, when we increase the SGWB self-noise, the sensitivity to single sources remains consistent across all observing strategies. 
 
As the MPTA aims to optimise for both single source and SGWB detection, we need to understand how the choice of single source strategy affects SGWB sensitivity. For the same observing strategies,  we also compared the SGWB sensitivity. At lower frequencies, case (b) and case (c) provide nearly identical sensitivities for low GW self-noise. Case (a) and case (b) also provide nearly identical sensitivities for GW self-noise $2 \times 10^{-15}$ and $5 \times 10^{-15}$. We found that case (c) is less sensitive as we increase the SGWB self-noise, while the other strategies decrease the sensitivity to an SGWB by only 2-3\%. This could be because the inter-pulsar correlations contribute significantly to the sensitivity of an SGWB rather than the high-cadence and high precision of the best pulsars. Therefore, out of the different observing strategies, case (b) optimises the sensitivity to single sources with minimal reduction in sensitivity to an SGWB while keeping the overall observing time constant.
We note that we have not considered the ability of the array to angularly resolve a source and have used a sky-averaged approximation only.
Angular resolution is improved with timing a larger number of pulsars, so there is likely a trade-off between timing fewer pulsars to higher precision and being able to localise an individual source. We defer this analysis to future work. 


\subsection{Challenges of timing with UHF-band}
The MPTA has observed MSPs exclusively using the L-band receiver. 
In contrast, most of the PTAs use more than one telescope and therefore, different backends, multiple receivers, or both, which makes it important for them to address the systematic differences during data combination. 
In our study, we showed that observing pulsars with the UHF receiver of MeerKAT could improve the sensitivity of the MPTA more significantly than other chosen observing strategies, under ideal assumptions.
While our assumptions include a variety of chromatic noise processes (both DM noise and chromatic noise potentially attributed to scattering), our modelling assumes they can be perfectly corrected.
It is possible that the DM (and hence DM variations) of a pulsar are frequency-dependent as the ray bundles at different frequencies trace different lines of sight in the IISM  \cite[][]{2016ApJ...817...16C, 2020ApJ...892...89L}. 
This has been predicted to cause the dispersion measure to vary at different frequencies, and has been tentatively detected in the PSR~J2219+4754 and PSR~J2241$-$5236 \cite[][]{2019A&A...624A..22D, 2022ApJ...930L..27K}. In a similar effect, multi-path propagation through non-uniform small-scale density fluctuations in the IISM causes the lower frequency emission to be highly scattered, resulting in a decorrelation in chromatic noise between different frequency components \cite[][]{2017MNRAS.464.2075S}. 
If such decorrelated scattering noise is present in our observations, it could reduce the sensitivity of low-frequency observations.
In many PTA experiments, band noise has been identified in pulsar timing data sets, modelled as additional noise unique to each observing frequency sub-band \cite[][]{2016MNRAS.458..868L,2023ApJ...951L...7R}. This could be an example of such propagation noise.

\begin{figure}
    \centering
    \subfloat{
        \includegraphics[width=0.95\columnwidth]{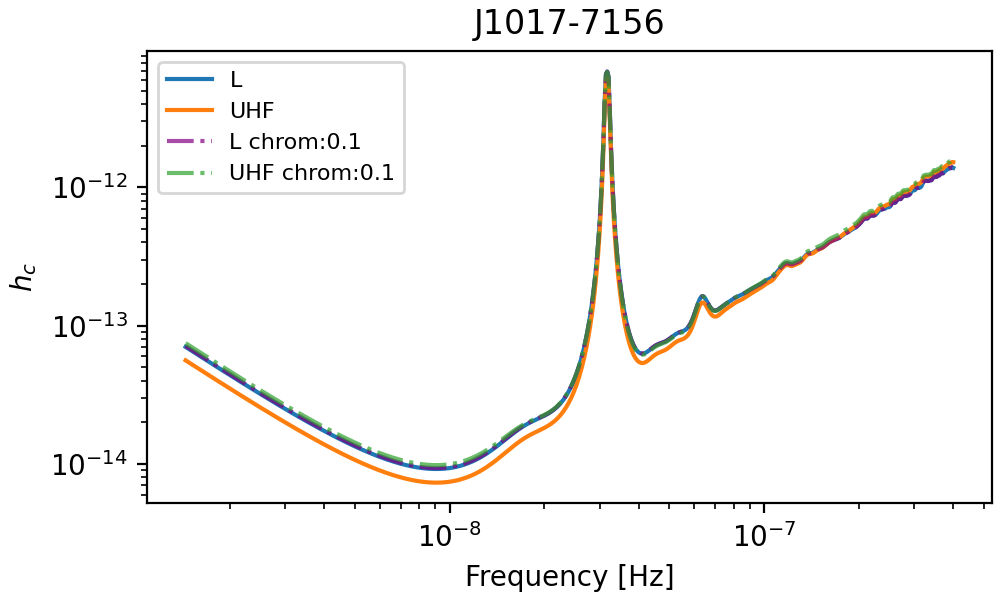}} 
    \vfill
    \subfloat{
        \includegraphics[width=0.95\columnwidth]{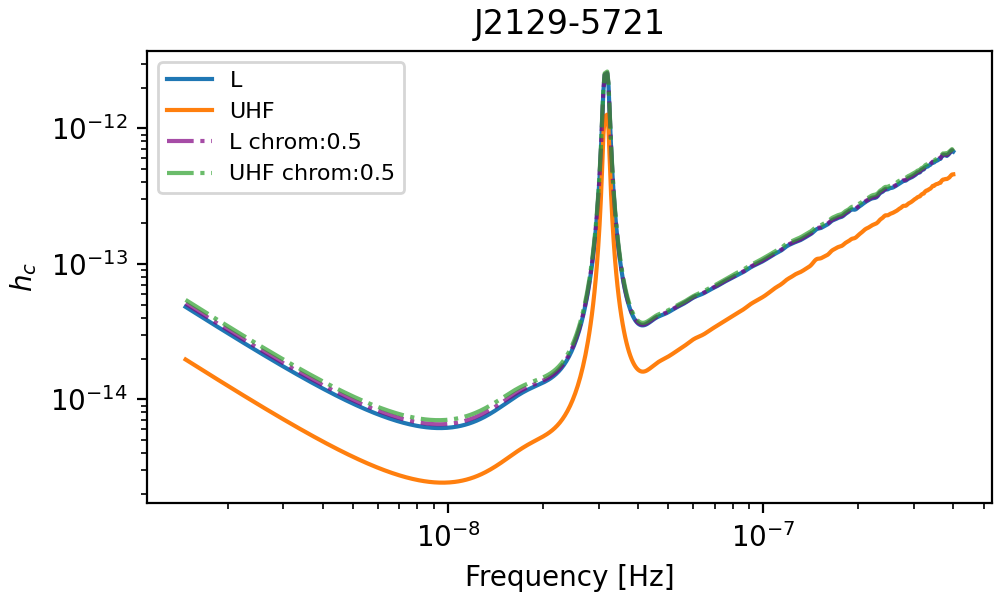}} 
    \caption{Sensitivity curves for individual pulsars with decorrelated chromatic noise. Top panel: For PSR~J1017-7156, UHF is less sensitive than the L-band with added chromatic noise at 10\% decorrelation. Bottom panel: In the case of PSR~J2129--5721, UHF is less sensitive than the L-band with added chromatic noise at 50\% decorrelation. The solid lines show the sensitivity curves using the fully correlated chromatic noise.}
    \label{fig:chrom}
\end{figure}

To assess the impact of such noise on the sensitivity of the MPTA, we developed a new model that includes partially correlated chromatic noise, similar to band noise modelling in \citet{2016ApJ...819..155L, 2021MNRAS.502..478G, 2023ApJ...951L...7R}. In this alternate chromatic noise model, one component of the noise is fully correlated in frequency and time, while another component is uncorrelated between different sub-bands (but has the same temporal correlation). 
The partially correlated noise was added as a power law, with power spectral density following equation \ref{eq:cn} for ToAs, selected by frequency sub-band. Since our aim was to determine what combination of partially correlated noise would alter our UHF sensitivity projections given in \ref{sec:uhf_subarray}, we modelled this additional noise assuming a range of fractional correlations. We varied the level of correlation by adjusting the amplitude of the power spectral density of the correlated components to assess the potential impact of such an effect.  
When we decrease the degree of correlation, the UHF-only observing strategy becomes less sensitive than the L-band. The level depends on the choice of the pulsar. Two of the most extreme cases are shown in Figure \ref{fig:chrom}. The dashed curves in Figure \ref{fig:chrom} show the final L-band and UHF-band sensitivities for individual pulsars at the degree of correlation where the UHF-band becomes less sensitive than the L-band. For PSR~J2129$-$5721, the UHF-band becomes less sensitive than the L-band if the fraction of chromatic noise that is correlated is 50\% whereas for many pulsars, even smaller levels (10\%) of correlated chromatic noise decrease UHF sensitivity, which is the case for PSR~J1017$-$7156. This new chromatic noise model is important for understanding the effects of decorrelated chromatic noise and gives a perspective on the diminishing low-frequency effects. Using dedicated observing campaigns at low frequencies, we can determine if such processes are present in PTA data sets.


\begin{figure*}
\centering
\includegraphics[width=0.8\textwidth]{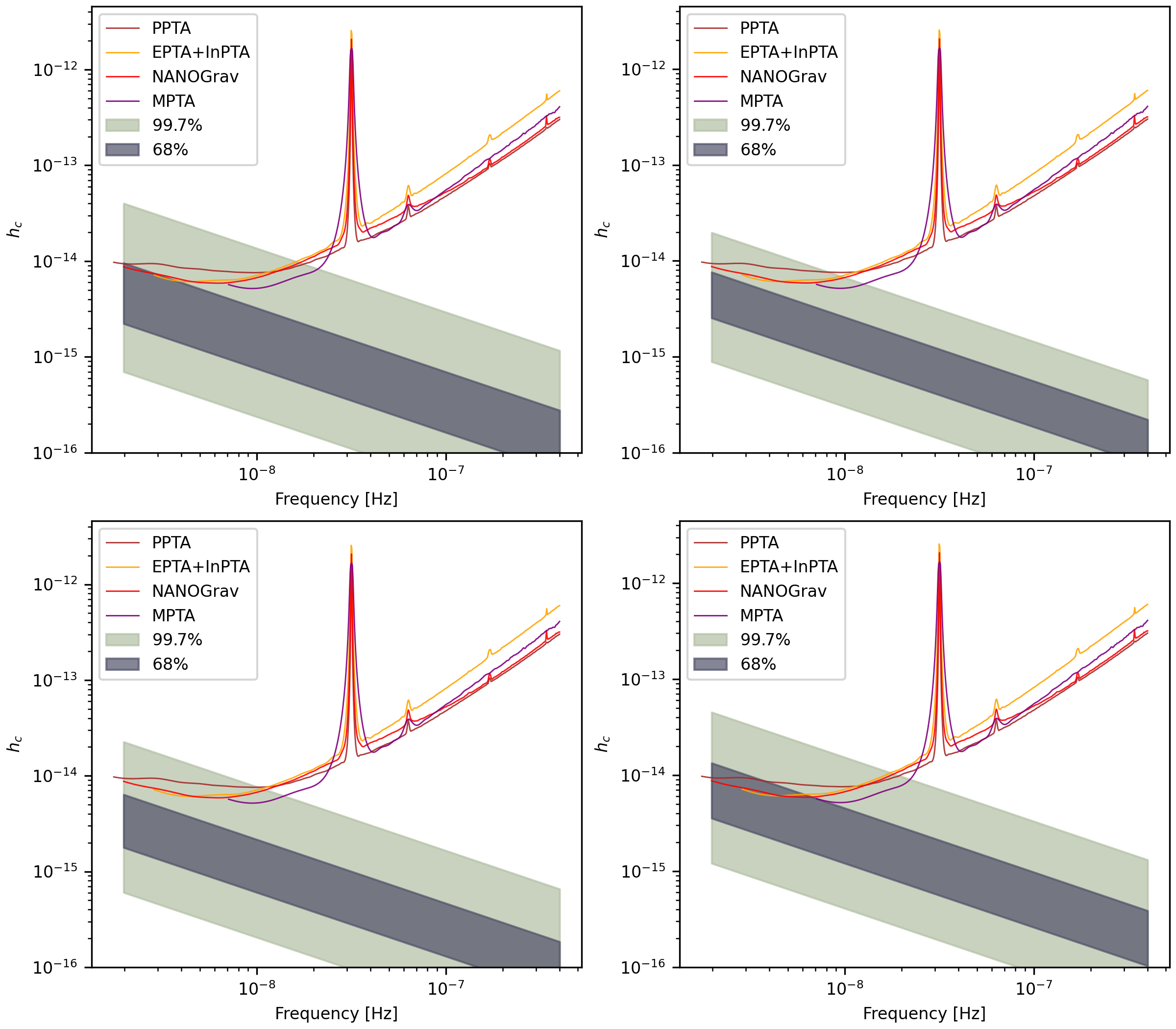}
\caption{Comparison of MPTA 4.5-year sensitivity with the other PTAs for an SGWB at an amplitude $2 \times 10^{-15}$. The shaded region shows the 68\% and 99.7\% confidence region of SGWB amplitudes derived from merger rates from redshift-dependent galaxy mass functions, fraction of close galaxy pairs and overmassive BHs in galaxies. The top-left panel includes all models described in \cite{2013MNRAS.433L...1S}, the top-right panel includes only the best estimates for models, the bottom-left panel includes models accounting for observations of overmassive BHs and the bottom-right panel includes models allowing for a broken power-law relation between stellar velocity dispersion and bulge stellar mass. }
\label{fig:mpta_sgwb}
\end{figure*}

\section{ Discussion}\label{sec:5}
\subsection{Comparison of PTA sensitivity}

The MPTA will continue to play an important role in international pulsar timing array efforts.
In Figure \ref{fig:mpta_sgwb}, we compare the sensitivity of the MPTA 4.5-year dataset with three major PTA collaborations, EPTA+InPTA, NANOGrav, and PPTA, assuming an SGWB amplitude of $2 \times 10^{-15}$. To create these curves, we have taken the dataset from the latest public data releases \citep{2023A&A...678A..48E, 2023ApJ...951L...9A, 2023PASA...40...49Z} and modelled the noise processes following the methodologies adopted by each PTA. The curves are calculated from a minimum GW frequency of $1/T$ where $T$ is the time span of the PTA data set. At higher frequencies between $\sim 10^{-8}- 10^{-7}$\,nHz, MPTA shows higher sensitivity than other PTAs, whereas at lower frequencies it is less sensitive because of its shorter data span. 

\begin{figure}
\centering
\includegraphics[width=\columnwidth]{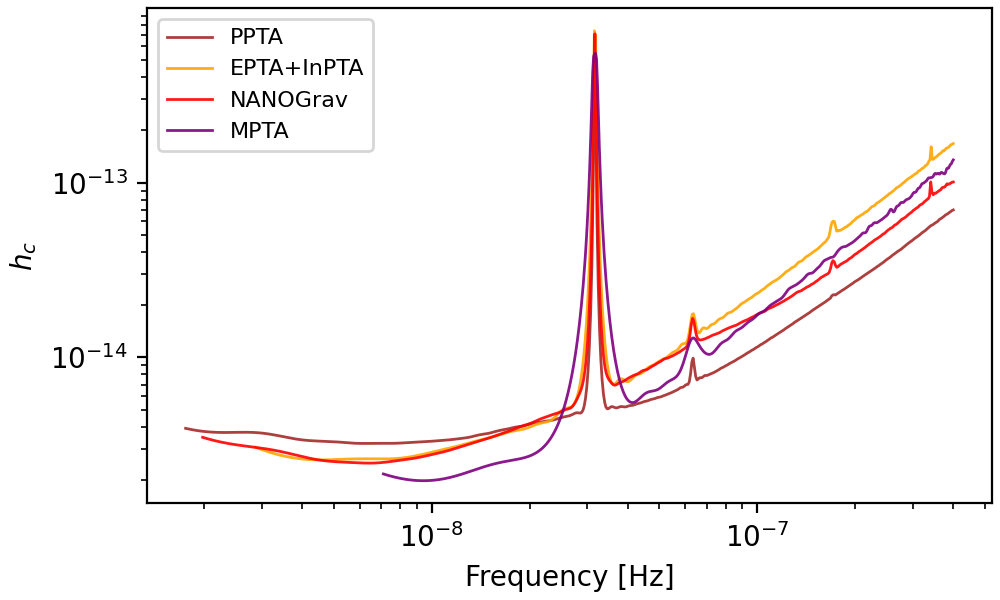}
\caption{ Comparison of MPTA sensitivity with other PTAs for a single source with an SGWB self-noise amplitude of $2 \times 10^{-15}$.}
\label{fig:mpta_cgwb}
\end{figure}

We have also compared the sensitivity of the MPTA to predictions for the amplitude of the SGWB in Figure \ref{fig:mpta_sgwb}.
The shaded region describes the 68\% and 99.7\% credible intervals for SGWB amplitudes derived using the four models presented in \cite{2013MNRAS.433L...1S}, based on different SMBH population and galaxy merger-rate studies. When combining all galaxy merger rates and galaxy pair count estimates (top left), a higher spread in the SGWB amplitudes is predicted.  The spread reduces if only the best estimates of galaxy mass functions and pair counts are used (top right). The bottom left panel shows the SGWB amplitude limits for models including galaxy mass functions that account for overmassive BHs, such as found in the centre of bright cluster galaxies \citep{2024ApJ...960L...1N} and in the bottom right panel, the SGWB amplitude limits account for a broken power law SMBH mass-galaxy bulge scaling relation. By improving the sensitivity of PTAs to weaker SGWB amplitudes, we can improve detection probability with even more conservative models. 
Additionally, we plotted the sky-averaged single-source sensitivity curves for the  MPTA and other major PTAs in Figure \ref{fig:mpta_cgwb}. The MPTA single-source sensitivity curve is lower than the SGWB sensitivity curve because the HD correlations reduce the number of pulsar pairs contributing effectively to the stochastic optimal statistic \citep{2019PhRvD.100j4028H}. The MPTA shows better sensitivity than the other PTAs at higher frequencies. Even though these sensitivities represent the sky-averaged metric, at these higher frequencies, the MPTA can leverage the targeted search for individual SMBHBs in the local universe ($\sim$50\,Mpc).

\subsection{S/N evolution with time}

\begin{figure}
\centering
\includegraphics[width=\columnwidth]{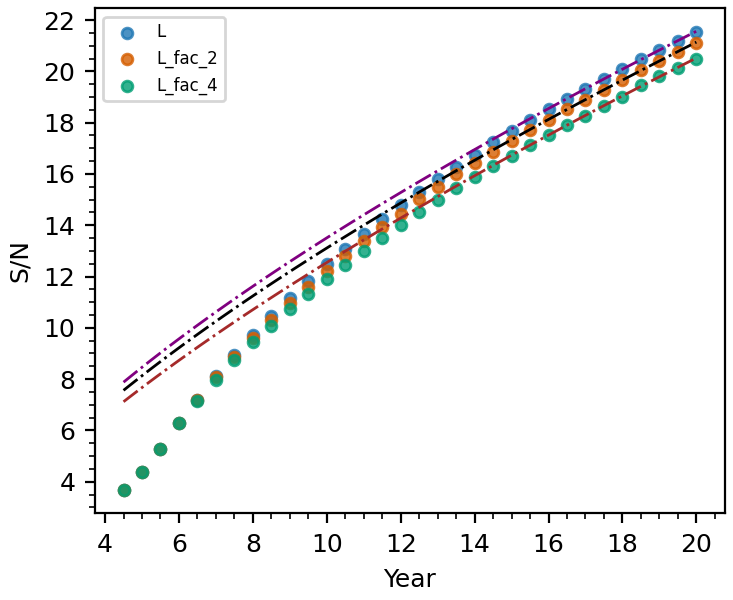}
\caption{ Evolution of S/N with time span of MPTA. We show the  S/N variations for simulated datasets up to 20 years for three observing strategies: standard L-band  (L), twice cadence observations (L\_fac\_2) and quadrupled cadence observations for best MSPs (L\_fac\_4).
The S/N is estimated for an SGWB self-noise amplitude at $5 \times 10^{-15}$, i.e. the CURN amplitude of MPTA. The dashed lines represent a power law model fit to the last four years of the dataset and indicate a transition from the weak signal regime towards a strong signal regime for the SGWB.}
\label{fig:mpta_yr}
\end{figure}

We can also forecast how the MPTA sensitivity will improve with time.
The sensitivity increase depends on the amplitude of the SGWB relative to other noise sources in the dataset.
A PTA can be either in the weak signal limit, where the white noise of the pulsars dominates the noise budget, a strong signal limit where GW power is dominant in low frequencies or an intermediate signal limit where low frequencies of background are above the white noise level but the highest frequencies are below it \citep{2013CQGra..30v4015S}. 
Assuming SGWB self-noise amplitude of $5 \times 10^{-15}$ (which is the amplitude of a CURN in the 4.5-year MPTA dataset), we calculated how the S/N increases with time span. To do so, we have simulated longer spanning datasets assuming identical cadence to the current MPTA observations, observed using the L-band receiver, using the flux density and $W_{\rm eff}$ relation as described in Section \ref{sec:uhf}. The evolution of the S/N is shown in Figure \ref{fig:mpta_yr}. We also model how S/N varies with time for the MPTA. Since there is a clear change in the power law with time span, we fitted for a power law model to S/N  ($\rm S/N = \rm S (\rm T/{\rm 1\,yr})^\alpha$ in the last 4 years and found the best-fit values at an amplitude of $\rm S =7.88(4)$ and $\alpha=0.670(4)$.  This spectral index is slightly larger than expected for the strong signal regime ($\alpha=0.5$), suggesting that not all of the pulsars in the array are in the strong signal limit, even after $20$\,years.

We also compared the standard L-band observing strategy with two alternate observing strategies. In the first, the 20 best MSPs were observed with twice the cadence, while the remaining were observed at the reduced cadence to maintain the same total observing time. In the second, the best $20$ pulsars were observed with a quadruple cadence, while the remaining were again observed at reduced cadence to maintain the same observing time. 
For these two observing strategies, we also fitted the last 4-year dataset and the best-fit amplitudes are  $S=7.56(4)$ and $7.12(8)$ and the indices are $\alpha=0.690(4)$ and $0.71(8)$, respectively. We see a modest reduction in sensitivity when we increase the cadence. We attribute this to lower sensitivity in SGWB correlations involving the lower-cadence pulsars.





\section{Conclusions}\label{sec:6}
We have analysed the sensitivity of PTAs by accounting for intrinsic pulsar noise, noise modelling techniques, and observing strategies. 
Appropriate modelling of noise processes such as jitter noise and DM noise for the next generation telescopes with improved sensitivity and higher bandwidth will enable high S/N detections of nHz-frequency GWs and identify and allow study of the sources of GW emission.
Using higher sensitivity observations from a more extensive MeerKAT data set, we measured jitter noise for 20 new MSPs. Measurement of the fraction of jitter-limited observations revealed that more than 50\% of observations of 11 MSPs are jitter-dominated, including the brightest MSP, J0437$-$4715, which is jitter-dominated in all observations. 
Using jitter noise measurements and model predictions, we found that jitter noise will be a limiting factor for the precision timing of the 29 best-timed pulsars with the future SKA-Mid telescope. 
The presence of jitter-dominated pulsars will motivate the use of sub-arrays for the SKA-Mid telescope. 
PTAs contribute to the broader efforts of IPTA GW search by combining the data from individual PTAs, which will include improved backends and high-gain telescopes. It is essential to collaborate on observing strategies between PTAs to minimise the number of observations that are jitter-dominated. 
One possibility is that PTAs with less sensitivity observe the more jitter-limited pulsars for longer integration times, whereas more sensitive telescopes focus on the fainter MSPs. 

We have investigated two methods for DM noise modelling and found that using a stationary GP improves the sensitivity of the array by 44\% compared with using a piece-wise constant model. However, there could be model covariance between DM noise, chromatic noise, and variations in the solar wind, as these stochastic processes also have a radio frequency dependence. It is also possible that DM variations are not physically well modelled by a GP with a power-law power spectral density because of discrete structures in the IISM. In such cases, the inferred properties for the GP could be biased. This would likely decrease PTA sensitivity to GWs, and the improvement we found could be overestimated. 
 
We then considered alternative observing strategies. We used our existing L-band observations to forecast timing precisions in higher and lower-frequency observing bands that have not been used regularly for precision timing.
Even though low-frequency observations with the UHF receiver of MeerKAT provide an improvement in sensitivity, propagation effects can potentially contribute excess error to the timing residuals that counteract this. We developed an alternative model that accounts for partially correlated chromatic noise, and such a noise model reduces the improvement in sensitivity obtained from UHF-band observations. For individual pulsars, we found that only 10--50\% partial correlations can reduce the sensitivity of UHF-band observations to that of L-band.  
Many radio telescopes will soon have upgraded wide-bandwidth receiver systems, which provide an opportunity to observe MSPs at lower frequencies that might be able to mitigate these effects.  However, if chromatic noise is fully correlated and our projections about low-frequency observations are accurate, the effects of DM and other frequency-dependent factors can be fully mitigated. This could provide a unique advantage for current low-frequency observatories such as uGMRT and LOFAR.  

PTAs aim to optimise the array towards both SGWB and single source detections. Our sensitivity studies using different observing strategies revealed that the improvement over the traditional observing strategy is dependent on the amplitude of SGWB. We conclude that high-cadence observations for the few best MSPs with reduced cadence observations for the remaining will optimise PTAs for single sources with a marginal loss in sensitivity to SGWB detection. Our work complements the time allocation optimisation study by \citet{2025MNRAS.540..603M}. Using a simplified PTA assumption, their study showed that small modifications in the time allocation of MSPs can boost the S/N of detected SGWB by approximately 20\% for the MPTA. The sensitivity analysis and lessons learned here can be implemented by any PTA. As the MPTA continues to take data and contribute to the IPTA datasets, our optimisation strategies will lead to an improved S/N of GW detection. Post detection, this will enable robust studies of the nHz-frequency GW sky.

\begin{acknowledgement}

The MeerKAT telescope
is operated by the South African Radio Astronomy Observatory,
which is a facility of the National Research Foundation, an agency
of the Department of Science and Innovation. PG acknowledges the support of a Swinburne University of Technology Postgraduate Stipend scholarship. PG, RMS, MB, MM and DJR acknowledge support through the Australian Research Council (ARC) Centre of Excellence
grant CE170100004 and CE230100016.
RMS acknowledges support through
ARC Future Fellowship FT190100155. KG and DC acknowledge continuing valuable support from the Max-Planck Society. KG acknowledges support from the International Max Planck Research School (IMPRS) for Astronomy and Astrophysics at the Universities of Bonn and Cologne.
This work used
the OzSTAR national facility at Swinburne University of Technology
and the pulsar portal maintained by ADACS at URL: https://pulsars.org.au. 
OzSTAR and the pulsar portal are funded by Swinburne University of Technology and the
National Collaborative Research Infrastructure Strategy (NCRIS). MeerTime observations used the PTUSE computing cluster. This cluster was funded in part by the Max-Planck-Institut für Radioastronomie (MPIfR) and the Max-Planck-Gesellschaft.
\end{acknowledgement}

\section*{Software}
This work has made use of the following open-source Python-based software:
\code{Matplotlib} \citep{Hunter:2007}, \code{Numpy} \citep{harris2020array}, \code{bilby} \citep{bilby_paper}, \code{enterprise} \citep{enterprise}, \code{PulsePortraiture} \citep{2016ascl.soft06013P}, \code{libstempo} \citep{2020ascl.soft02017V} and \code{hasasia} \citep{2019PhRvD.100j4028H}.
\section*{Data Availability}
The dataset used in this work can be made available upon request. The dataset primarily comprises of the MPTA 4.5-year data release and can be found on the https://pulsars.org.au portal.

\begin{table*}
\centering

\caption{Measured jitter values and upper limits for 89 pulsars in the MPTA. For each pulsar we present the period of the pulsar (P), DM, weighted RMS of the timing residuals for the chosen observation, jitter noise in 1\,hr $\sigma_{\rm J}(\rm hr)$, jitter noise for a single pulse $\sigma_{\rm J,1}$ and comparison of $\sigma_{\rm J}(\rm hr)$ with previous studies. S14, L16, L19, P21 and W24 refer to \citet{2014MNRAS.443.1463S}, \citet{2016ApJ...819..155L}, \citet{2019ApJ...872..193L}, \citet{2021MNRAS.502..407P} and \citet{2024ApJ...964....6W} respectively.}
\label{tab:jitter}
\begin{tabular}{rrrrrrc}
\hline
PSRNAME    & Period & DM & RMS    & $\sigma_{\rm J}(\rm hr)$ & $\sigma_{\rm J}(1)$ & Previous work\\
& (ms) & (pc/cc) & & (ns) & ($\mu$s) &\\
\hline
J0030+0451 & 4.9 & 4.3 & 1.268 & <78 & & <60(P21), 61.1±3.5(L19), 153(L16), 46±1(W24) \\ 
J0101$-$6422 & 2.6 & 11.9 & 1.427 & <72 & &    \\ 
J0125$-$2327 & 3.7 & 9.6 & 0.635 & 26$\pm$3 & 26$\pm$3 & 48±13(P21)   \\ 
J0437$-$4715 & 5.8 & 2.6 & 1.108 & 52$\pm$7 & 41$\pm$6 & 50±10(P21), 48.0±0.6(S14)   \\ 
J0610$-$2100 & 3.9 & 60.7 & 1.758 & <110 & &    \\ 
J0613$-$0200 & 3.1 & 38.8 & 0.85 & <40 & & <400(S14), 133$\pm$8(L19), 34.4±0.2(W24)   \\ 
J0614$-$3329 & 3.1 & 37.1 & 1.121 & <64 & &    \\ 
J0636$-$3044 & 3.9 & 15.5 & 3.051 & 88$\pm$5 & 84$\pm$5 & 100±30(P21)   \\ 
J0711$-$6830 & 5.5 & 18.4 & 0.931 & 40$\pm$5 & 32$\pm$4 & 60±20(P21), <90(S14)   \\ 
J0900$-$3144 & 11.1 & 75.7 & 2.965 & 83$\pm$11 & 47$\pm$6 & <130(P21)   \\ 
J0931$-$1902 & 4.6 & 41.5 & 2.038 & <120 & &    \\ 
J0955$-$6150 & 2.0 & 160.9 & 2.606 & <156 & &    \\ 
J1012$-$4235 & 3.1 & 71.7 & 2.066 & <131 & &    \\ 
J1017$-$7156 & 2.3 & 94.2 & 0.194 & <9 & & <10(P21), <100(S14)   \\ 
J1022+1001 & 16.5 & 10.3 & 2.267 & 106$\pm$14 & 50$\pm$7 & 120±20(P21), 280±140(S14), 265±20(L19)  \\ 
J1024$-$0719 & 5.2 & 6.5 & 0.757 & 30$\pm$4 & 25$\pm$3 & <30(P21), 18±10(L19)   \\ 
J1036$-$8317 & 3.4 & 27.1 & 1.073 & <67 & &    \\ 
J1045$-$4509 & 7.5 & 58.1 & 4.451 & 185$\pm$24 & 128$\pm$17 & 130±75(P21), <900(S14)   \\ 
J1101$-$6424 & 5.1 & 207.4 & 7.665 & <410 & &    \\ 
J1103$-$5403 & 3.4 & 103.9 & 1.372 & <66 & &    \\ 
J1125$-$5825 & 3.1 & 124.8 & 2.256 & <151 & &    \\ 
J1125$-$6014 & 2.6 & 52.9 & 0.435 & <22 & &    \\ 
J1216$-$6410 & 3.5 & 47.4 & 0.854 & <50 & &    \\ 
J1231$-$1411 & 3.7 & 8.1 & 1.347 & 38$\pm$2 & 38$\pm$2 &    \\ 
J1327$-$0755 & 2.7 & 27.9 & 0.926 & <44 & &    \\ 
J1421$-$4409 & 6.4 & 54.6 & 5.714 & <302 & &    \\ 
J1431$-$5740 & 4.1 & 131.4 & 3.032 & 92$\pm$11 & 86$\pm$10 &    \\ 
J1435$-$6100 & 9.3 & 113.8 & 3.704 & <175 & &    \\ 
J1446$-$4701 & 2.2 & 55.8 & 0.999 & <48 & &    \\ 
J1455$-$3330 & 8.0 & 13.6 & 2.101 & 81$\pm$8 & 54$\pm$5 &   150$\pm$18(L19) \\ 
J1513$-$2550 & 2.1 & 46.9 & 17.006 & 788$\pm$97 & 1027$\pm$126 &    \\ 
J1514$-$4946 & 3.6 & 31.0 & 0.987 & <57 & &    \\ 
J1525$-$5545 & 11.4 & 127.0 & 7.028 & <387 & &    \\ 
J1543$-$5149 & 2.1 & 51.0 & 2.029 & <110 & &    \\ 
J1545$-$4550 & 3.6 & 68.4 & 0.996 & <56 & &    \\ 
J1547$-$5709 & 4.3 & 95.7 & 6.491 & 171$\pm$14 & 157$\pm$13 &    \\ 
J1600$-$3053 & 3.6 & 52.3 & 0.542 & 19$\pm$2 & 19$\pm$2 & <30(P21), <200(S14)   \\ 
J1603$-$7202 & 14.8 & 38.0 & 2.99 & 136$\pm$18 & 67$\pm$9 & 180±40(P21), 300±56(S14)   \\ 
J1614$-$2230 & 3.2 & 34.5 & 0.759 & <41 & & 60$\pm$7(L19)   \\ 
J1628$-$3205 & 3.2 & 42.1 & 5.143 & <244 & &    \\ 
J1629$-$6902 & 6.0 & 29.5 & 1.103 & 31$\pm$4 & 24$\pm$3 & <60(P21)   \\ 
J1643$-$1224 & 4.6 & 62.4 & 1.141 & <59 & & <60(P21), <500(S14), 31±12(L19), 40.8±0.3(W24) \\ 
J1652$-$4838 & 3.8 & 188.2 & 4.489 & <224 & &    \\ 
J1653$-$2054 & 4.1 & 56.5 & 5.177 & <257 & &    \\ 
J1658$-$5324 & 2.4 & 30.8 & 2.254 & <124 & &    \\ 
J1705$-$1903 & 2.5 & 57.5 & 0.283 & <15 & &    \\ 
J1708$-$3506 & 4.5 & 146.8 & 10.889 & <519 & &    \\ 
J1713+0747 & 4.6 & 16.0 & 0.586 & 27$\pm$3 & 24$\pm$3 & 36.0$\pm$0.1(L19), 24.9±0.3(W24)   \\ 

\hline

\end{tabular}
\end{table*}

\begin{table*}
\centering
\begin{tabular}{rrrrrrc}
\hline
PSRNAME    & Period & DM & RMS    & $\sigma_{\rm J}(\rm hr)$ & $\sigma_{\rm J,1}$ & Previous work\\
& (ms) & (pc/cc) & & (ns) & ($\mu$s) &\\
\hline
J1719$-$1438 & 5.8 & 36.8 & 3.962 & 96$\pm$7 & 76$\pm$6 &    \\ 
J1721$-$2457 & 3.5 & 48.2 & 8.843 & <440 & &    \\ 
J1730$-$2304 & 8.1 & 9.6 & 2.391 & 109$\pm$14 & 73$\pm$9 & 80±45(P21), <400(S14)   \\ 
J1731$-$1847 & 2.3 & 106.5 & 1.272 & <82 & &    \\ 
J1732$-$5049 & 5.3 & 56.8 & 0.921 & <44 & &    \\ 
J1737$-$0811 & 4.2 & 55.3 & 5.523 & <284 & &    \\ 
J1744$-$1134 & 4.1 & 3.1 & 0.79 & 37$\pm$5 & 35$\pm$5 & 30±6(P21), 37.8±0.8(S14), 44±1(L19), 29.4±0.2(W24) \\ 
J1747$-$4036 & 1.6 & 152.9 & 2.943 & 85$\pm$11 & 126$\pm$16 &    \\ 
J1751$-$2857 & 3.9 & 42.8 & 4.224 & <239 & &    \\ 
J1756$-$2251 & 28.5 & 121.2 & 8.242 & 256$\pm$15 & 91$\pm$5 & <500(P21)   \\ 
J1757$-$5322 & 8.9 & 30.8 & 2.74 & 87$\pm$9 & 55$\pm$6 & 130±45(P21)   \\ 
J1801$-$1417 & 3.6 & 57.3 & 1.924 & <112 & &    \\ 
J1802$-$2124 & 12.6 & 149.6 & 1.969 & <100 & & <80(P21)   \\ 
J1804$-$2717 & 9.3 & 24.7 & 2.295 & 72$\pm$9 & 45$\pm$6 &    \\ 

J1804$-$2858 & 1.5 & 232.5 & 10.335 & <529 & &    \\ 
J1811$-$2405 & 2.7 & 60.6 & 1.07 & <59 & &    \\ 
J1825$-$0319 & 4.6 & 119.6 & 5.647 & <316 & &    \\ 
J1832$-$0836 & 2.7 & 28.2 & 1.076 & 37$\pm$5 & 43$\pm$6 &  <49(L19)  \\ 
J1843$-$1113 & 1.8 & 60.0 & 0.57 & 16$\pm$2 & 22$\pm$3 &    \\ 
J1843$-$1448 & 5.5 & 114.5 & 36.929 & <1759 & &    \\ 
J1902$-$5105 & 1.7 & 36.3 & 1.216 & 34$\pm$4 & 49$\pm$6 &    \\ 
J1903$-$7051 & 3.6 & 19.7 & 1.054 & 30$\pm$4 & 30$\pm$4 &    \\ 
J1909$-$3744 & 2.9 & 10.4 & 0.219 & 10$\pm$1 & 11$\pm$1 & 9±3(P21), 8.6±0.8(S14), 14±0.5(L19),  \\ 
J1911$-$1114 & 3.6 & 31.0 & 2.006 & <105 & &    \\ 
J1918$-$0642 & 7.6 & 26.6 & 1.422 & 49$\pm$6 & 34$\pm$4 & <55(P21), 53$\pm$9(L19), 55.7±0.5(W24)   \\ 
J1933$-$6211 & 3.5 & 11.5 & 0.946 & 31$\pm$4 & 31$\pm$4 &    \\ 
J1946$-$5403 & 2.7 & 23.7 & 0.269 & <16 & & <9(P21)   \\ 
J2010$-$1323 & 5.2 & 22.2 & 1.089 & <66 & & <80(P21), 59±3(L19)   \\ 
J2039$-$3616 & 3.3 & 24.0 & 0.723 & <43 & & <25(P21)   \\ 
J2124$-$3358 & 4.9 & 4.6 & 1.128 & 42$\pm$5 & 36$\pm$4 &    \\ 
J2129$-$5721 & 3.7 & 31.8 & 0.456 & <22 & & <11(P21), <400(S14)   \\ 
J2145$-$0750 & 16.1 & 9.0 & 3.525 & 165$\pm$21 & 78$\pm$10 & 200±20(P21), 192±6(S14), 173±4(L19), 168±1(W24) \\ 
J2150$-$0326 & 3.5 & 20.7 & 1.434 & 52$\pm$7 & 53$\pm$7 &    \\ 
J2222$-$0137 & 32.8 & 3.3 & 4.398 & 204$\pm$13 & 68$\pm$4 &    \\ 
J2229+2643 & 3.0 & 22.7 & 0.989 & 31$\pm$2 & 34$\pm$2 &  81$\pm$11(L19)  \\ 
J2234+0944 & 3.6 & 17.8 & 0.464 & 15$\pm$2 & 15$\pm$2 &   40$\pm$2(L19) \\ 
J2236$-$5527 & 6.9 & 20.1 & 2.085 & 79$\pm$3 & 57$\pm$2 &    \\ 
J2241$-$5236 & 2.2 & 11.4 & 0.088 & 3.5$\pm$0.3 & 4.5$\pm$0.4 & 3.8±0.8 (P21), <50(S14)   \\ 
J2317+1439 & 3.4 & 21.9 & 0.575 & 18$\pm$2 & 18$\pm$2 &   24$\pm$2(L19) \\ 
J2322+2057 & 4.8 & 13.4 & 1.693 & 40$\pm$3 & 35$\pm$3 &  31$\pm$11(L19)  \\ 
J2322$-$2650 & 3.5 & 6.1 & 1.403 & <68 & &    \\ 
\hline
\end{tabular}
\end{table*}

\bibliography{example_new}

\begin{thebibliography}{}
\expandafter\ifx\csname natexlab\endcsname\relax\def\natexlab#1{#1}\fi

\bibitem[{{Agazie} {et~al.}(2023{\natexlab{a}}){Agazie}, {Anumarlapudi}, {Archibald}, {Arzoumanian}, {Baker}, {B{\'e}csy}, {Blecha}, {Brazier}, {Brook}, {Burke-Spolaor}, {Burnette}, {Case}, {Charisi}, {Chatterjee}, {Chatziioannou}, {Cheeseboro}, {Chen}, {Cohen}, {Cordes}, {Cornish}, {Crawford}, {Cromartie}, {Crowter}, {Cutler}, {Decesar}, {Degan}, {Demorest}, {Deng}, {Dolch}, {Drachler}, {Ellis}, {Ferrara}, {Fiore}, {Fonseca}, {Freedman}, {Garver-Daniels}, {Gentile}, {Gersbach}, {Glaser}, {Good}, {G{\"u}ltekin}, {Hazboun}, {Hourihane}, {Islo}, {Jennings}, {Johnson}, {Jones}, {Kaiser}, {Kaplan}, {Kelley}, {Kerr}, {Key}, {Klein}, {Laal}, {Lam}, {Lamb}, {Lazio}, {Lewandowska}, {Littenberg}, {Liu}, {Lommen}, {Lorimer}, {Luo}, {Lynch}, {Ma}, {Madison}, {Mattson}, {McEwen}, {McKee}, {McLaughlin}, {McMann}, {Meyers}, {Meyers}, {Mingarelli}, {Mitridate}, {Natarajan}, {Ng}, {Nice}, {Ocker}, {Olum}, {Pennucci}, {Perera}, {Petrov}, {Pol}, {Radovan}, {Ransom}, {Ray}, {Romano}, {Sardesai}, {Schmiedekamp}, {Schmiedekamp},
  {Schmitz}, {Schult}, {Shapiro-Albert}, {Siemens}, {Simon}, {Siwek}, {Stairs}, {Stinebring}, {Stovall}, {Sun}, {Susobhanan}, {Swiggum}, {Taylor}, {Taylor}, {Turner}, {Unal}, {Vallisneri}, {van Haasteren}, {Vigeland}, {Wahl}, {Wang}, {Witt}, {Young}, \& {Nanograv Collaboration}}]{2023ApJ...951L...8A}
{Agazie}, G., {Anumarlapudi}, A., {Archibald}, A.~M., {et~al.} 2023{\natexlab{a}}, \apjl, 951, L8

\bibitem[{{Agazie} {et~al.}(2023{\natexlab{b}}){Agazie}, {Alam}, {Anumarlapudi}, {Archibald}, {Arzoumanian}, {Baker}, {Blecha}, {Bonidie}, {Brazier}, {Brook}, {Burke-Spolaor}, {B{\'e}csy}, {Chapman}, {Charisi}, {Chatterjee}, {Cohen}, {Cordes}, {Cornish}, {Crawford}, {Cromartie}, {Crowter}, {Decesar}, {Demorest}, {Dolch}, {Drachler}, {Ferrara}, {Fiore}, {Fonseca}, {Freedman}, {Garver-Daniels}, {Gentile}, {Glaser}, {Good}, {G{\"u}ltekin}, {Hazboun}, {Jennings}, {Jessup}, {Johnson}, {Jones}, {Kaiser}, {Kaplan}, {Kelley}, {Kerr}, {Key}, {Kuske}, {Laal}, {Lam}, {Lamb}, {Lazio}, {Lewandowska}, {Lin}, {Liu}, {Lorimer}, {Luo}, {Lynch}, {Ma}, {Madison}, {Maraccini}, {McEwen}, {McKee}, {McLaughlin}, {McMann}, {Meyers}, {Mingarelli}, {Mitridate}, {Ng}, {Nice}, {Ocker}, {Olum}, {Panciu}, {Pennucci}, {Perera}, {Pol}, {Radovan}, {Ransom}, {Ray}, {Romano}, {Salo}, {Sardesai}, {Schmiedekamp}, {Schmiedekamp}, {Schmitz}, {Shapiro-Albert}, {Siemens}, {Simon}, {Siwek}, {Stairs}, {Stinebring}, {Stovall}, {Susobhanan}, {Swiggum},
  {Taylor}, {Turner}, {Unal}, {Vallisneri}, {Vigeland}, {Wahl}, {Wang}, {Witt}, {Young}, \& {Nanograv Collaboration}}]{2023ApJ...951L...9A}
{Agazie}, G., {Alam}, M.~F., {Anumarlapudi}, A., {et~al.} 2023{\natexlab{b}}, \apjl, 951, L9

\bibitem[{{Agazie} {et~al.}(2024){Agazie}, {Antoniadis}, {Anumarlapudi}, {Archibald}, {Arumugam}, {Arumugam}, {Arzoumanian}, {Askew}, {Babak}, {Bagchi}, {Bailes}, {Bak Nielsen}, {Baker}, {Bassa}, {Bathula}, {B{\'e}csy}, {Berthereau}, {Bhat}, {Blecha}, {Bonetti}, {Bortolas}, {Brazier}, {Brook}, {Burgay}, {Burke-Spolaor}, {Burnette}, {Caballero}, {Cameron}, {Case}, {Chalumeau}, {Champion}, {Chanlaridis}, {Charisi}, {Chatterjee}, {Chatziioannou}, {Cheeseboro}, {Chen}, {Chen}, {Cognard}, {Cohen}, {Coles}, {Cordes}, {Cornish}, {Crawford}, {Cromartie}, {Crowter}, {Cury{\l}o}, {Cutler}, {Dai}, {Dandapat}, {Deb}, {DeCesar}, {DeGan}, {Demorest}, {Deng}, {Desai}, {Desvignes}, {Dey}, {Dhanda-Batra}, {Di Marco}, {Dolch}, {Drachler}, {Dwivedi}, {Ellis}, {Falxa}, {Feng}, {Ferdman}, {Ferrara}, {Fiore}, {Fonseca}, {Franchini}, {Freedman}, {Gair}, {Garver-Daniels}, {Gentile}, {Gersbach}, {Glaser}, {Good}, {Goncharov}, {Gopakumar}, {Graikou}, {Griessmeier}, {Guillemot}, {G{\"u}ltekin}, {Guo}, {Gupta}, {Grunthal}, {Hazboun},
  {Hisano}, {Hobbs}, {Hourihane}, {Hu}, {Iraci}, {Islo}, {Izquierdo-Villalba}, {Jang}, {Jawor}, {Janssen}, {Jennings}, {Jessner}, {Johnson}, {Jones}, {Joshi}, {Kaiser}, {Kaplan}, {Kapur}, {Kareem}, {Karuppusamy}, {Keane}, {Keith}, {Kelley}, {Kerr}, {Key}, {Kharbanda}, {Kikunaga}, {Klein}, {Kolhe}, {Kramer}, {Krishnakumar}, {Kulkarni}, {Laal}, {Lackeos}, {Lam}, {Lamb}, {Larsen}, {Lazio}, {Lee}, {Levin}, {Lewandowska}, {Littenberg}, {Liu}, {Liu}, {Liu}, {Lommen}, {Lorimer}, {Lower}, {Luo}, {Luo}, {Lynch}, {Lyne}, {Ma}, {Maan}, {Madison}, {Main}, {Manchester}, {Mandow}, {Mattson}, {McEwen}, {McKee}, {McLaughlin}, {McMann}, {Meyers}, {Meyers}, {Mickaliger}, {Miles}, {Mingarelli}, {Mitridate}, {Natarajan}, {Nathan}, {Ng}, {Nice}, {Ni{\c{t}}u}, {Nobleson}, {Ocker}, {Olum}, {Os{\l}owski}, {Paladi}, {Parthasarathy}, {Pennucci}, {Perera}, {Perrodin}, {Petiteau}, {Petrov}, {Pol}, {Porayko}, {Possenti}, {Prabu}, {Quelquejay Leclere}, {Radovan}, {Rana}, {Ransom}, {Ray}, {Reardon}, {Rogers}, {Romano}, {Russell},
  {Samajdar}, {Sanidas}, {Sardesai}, {Schmiedekamp}, {Schmiedekamp}, {Schmitz}, {Schult}, {Sesana}, {Shaifullah}, {Shannon}, {Shapiro-Albert}, {Siemens}, {Simon}, {Singha}, {Siwek}, {Speri}, {Spiewak}, {Srivastava}, {Stairs}, {Stappers}, {Stinebring}, {Stovall}, {Sun}, {Surnis}, {Susarla}, {Susobhanan}, {Swiggum}, {Takahashi}, {Tarafdar}, {Taylor}, {Taylor}, {Theureau}, {Thrane}, {Thyagarajan}, {Tiburzi}, {Toomey}, {Turner}, {Unal}, {Vallisneri}, {van der Wateren}, {van Haasteren}, {Vecchio}, {Venkatraman Krishnan}, {Verbiest}, {Vigeland}, {Wahl}, {Wang}, {Wang}, {Witt}, {Wang}, {Wang}, {Wayt}, {Wu}, {Young}, {Zhang}, {Zhang}, {Zhu}, {Zic}, \& {International Pulsar Timing Array Collaboration}}]{2024ApJ...966..105A}
{Agazie}, G., {Antoniadis}, J., {Anumarlapudi}, A., {et~al.} 2024, \apj, 966, 105

\bibitem[{{Allen} {et~al.}(2023){Allen}, {Dhurandhar}, {Gupta}, {McLaughlin}, {Natarajan}, {Shannon}, {Thrane}, \& {Vecchio}}]{2023arXiv230404767A}
{Allen}, B., {Dhurandhar}, S., {Gupta}, Y., {et~al.} 2023, arXiv e-prints, arXiv:2304.04767

\bibitem[{Ashton {et~al.}(2019)}]{bilby_paper}
Ashton, G., {et~al.} 2019, Astrophys. J. Suppl., 241, 27

\bibitem[{{Backer} {et~al.}(1982){Backer}, {Kulkarni}, {Heiles}, {Davis}, \& {Goss}}]{1982Natur.300..615B}
{Backer}, D.~C., {Kulkarni}, S.~R., {Heiles}, C., {Davis}, M.~M., \& {Goss}, W.~M. 1982, \nat, 300, 615

\bibitem[{{Baier} {et~al.}(2024){Baier}, {Hazboun}, \& {Romano}}]{2024arXiv240900336B}
{Baier}, J.~G., {Hazboun}, J.~S., \& {Romano}, J.~D. 2024, arXiv e-prints, arXiv:2409.00336

\bibitem[{{Bailes} {et~al.}(2020){Bailes}, {Jameson}, {Abbate}, {Barr}, {Bhat}, {Bondonneau}, {Burgay}, {Buchner}, {Camilo}, {Champion}, {Cognard}, {Demorest}, {Freire}, {Gautam}, {Geyer}, {Griessmeier}, {Guillemot}, {Hu}, {Jankowski}, {Johnston}, {Karastergiou}, {Karuppusamy}, {Kaur}, {Keith}, {Kramer}, {van Leeuwen}, {Lower}, {Maan}, {McLaughlin}, {Meyers}, {Os{\l}owski}, {Oswald}, {Parthasarathy}, {Pennucci}, {Posselt}, {Possenti}, {Ransom}, {Reardon}, {Ridolfi}, {Schollar}, {Serylak}, {Shaifullah}, {Shamohammadi}, {Shannon}, {Sobey}, {Song}, {Spiewak}, {Stairs}, {Stappers}, {van Straten}, {Szary}, {Theureau}, {Venkatraman Krishnan}, {Weltevrede}, {Wex}, {Abbott}, {Adams}, {Burger}, {Gamatham}, {Gouws}, {Horn}, {Hugo}, {Joubert}, {Manley}, {McAlpine}, {Passmoor}, {Peens-Hough}, {Ramudzuli}, {Rust}, {Salie}, {Schwardt}, {Siebrits}, {Van Tonder}, {Van Tonder}, \& {Welz}}]{2020PASA...37...28B}
{Bailes}, M., {Jameson}, A., {Abbate}, F., {et~al.} 2020, \pasa, 37, e028

\bibitem[{{Braun} {et~al.}(2019){Braun}, {Bonaldi}, {Bourke}, {Keane}, \& {Wagg}}]{2019arXiv191212699B}
{Braun}, R., {Bonaldi}, A., {Bourke}, T., {Keane}, E., \& {Wagg}, J. 2019, arXiv e-prints, arXiv:1912.12699

\bibitem[{{Burt} {et~al.}(2011){Burt}, {Lommen}, \& {Finn}}]{2011ApJ...730...17B}
{Burt}, B.~J., {Lommen}, A.~N., \& {Finn}, L.~S. 2011, \apj, 730, 17

\bibitem[{{Coles} {et~al.}(2010){Coles}, {Rickett}, {Gao}, {Hobbs}, \& {Verbiest}}]{2010ApJ...717.1206C}
{Coles}, W.~A., {Rickett}, B.~J., {Gao}, J.~J., {Hobbs}, G., \& {Verbiest}, J.~P.~W. 2010, \apj, 717, 1206

\bibitem[{{Coles} {et~al.}(2015){Coles}, {Kerr}, {Shannon}, {Hobbs}, {Manchester}, {You}, {Bailes}, {Bhat}, {Burke-Spolaor}, {Dai}, {Keith}, {Levin}, {Os{\l}owski}, {Ravi}, {Reardon}, {Toomey}, {van Straten}, {Wang}, {Wen}, \& {Zhu}}]{2015ApJ...808..113C}
{Coles}, W.~A., {Kerr}, M., {Shannon}, R.~M., {et~al.} 2015, \apj, 808, 113

\bibitem[{{Cordes} \& {Shannon}(2010)}]{2010arXiv1010.3785C}
{Cordes}, J.~M., \& {Shannon}, R.~M. 2010, arXiv e-prints, arXiv:1010.3785

\bibitem[{{Cordes} {et~al.}(2016){Cordes}, {Shannon}, \& {Stinebring}}]{2016ApJ...817...16C}
{Cordes}, J.~M., {Shannon}, R.~M., \& {Stinebring}, D.~R. 2016, \apj, 817, 16

\bibitem[{{Demorest} {et~al.}(2013){Demorest}, {Ferdman}, {Gonzalez}, {Nice}, {Ransom}, {Stairs}, {Arzoumanian}, {Brazier}, {Burke-Spolaor}, {Chamberlin}, {Cordes}, {Ellis}, {Finn}, {Freire}, {Giampanis}, {Jenet}, {Kaspi}, {Lazio}, {Lommen}, {McLaughlin}, {Palliyaguru}, {Perrodin}, {Shannon}, {Siemens}, {Stinebring}, {Swiggum}, \& {Zhu}}]{2013ApJ...762...94D}
{Demorest}, P.~B., {Ferdman}, R.~D., {Gonzalez}, M.~E., {et~al.} 2013, \apj, 762, 94

\bibitem[{{Donner} {et~al.}(2019){Donner}, {Verbiest}, {Tiburzi}, {Os{\l}owski}, {Michilli}, {Serylak}, {Anderson}, {Horneffer}, {Kramer}, {Grie{\ss}meier}, {K{\"u}nsem{\"o}ller}, {Hessels}, {Hoeft}, \& {Miskolczi}}]{2019A&A...624A..22D}
{Donner}, J.~Y., {Verbiest}, J.~P.~W., {Tiburzi}, C., {et~al.} 2019, \aap, 624, A22

\bibitem[{Ellis {et~al.}(2020)Ellis, Vallisneri, Taylor, \& Baker}]{enterprise}
Ellis, J.~A., Vallisneri, M., Taylor, S.~R., \& Baker, P.~T. 2020, ENTERPRISE: Enhanced Numerical Toolbox Enabling a Robust PulsaR Inference SuitE, Zenodo, doi:10.5281/zenodo.4059815

\bibitem[{{EPTA and InPTA Collaboration} {et~al.}(2023){EPTA and InPTA Collaboration}, {Antoniadis}, {Babak}, {Bak Nielsen}, {Bassa}, {Berthereau}, {Bonetti}, {Bortolas}, {Brook}, {Burgay}, {Caballero}, {Chalumeau}, {Champion}, {Chanlaridis}, {Chen}, {Cognard}, {Desvignes}, {Falxa}, {Ferdman}, {Franchini}, {Gair}, {Goncharov}, {Graikou}, {Grie{\ss}meier}, {Guillemot}, {Guo}, {Hu}, {Iraci}, {Izquierdo-Villalba}, {Jang}, {Jawor}, {Janssen}, {Jessner}, {Karuppusamy}, {Keane}, {Keith}, {Kramer}, {Krishnakumar}, {Lackeos}, {Lee}, {Liu}, {Liu}, {Lyne}, {McKee}, {Main}, {Mickaliger}, {Ni{\c{t}}u}, {Parthasarathy}, {Perera}, {Perrodin}, {Petiteau}, {Porayko}, {Possenti}, {Quelquejay Leclere}, {Samajdar}, {Sanidas}, {Sesana}, {Shaifullah}, {Speri}, {Spiewak}, {Stappers}, {Susarla}, {Theureau}, {Tiburzi}, {van der Wateren}, {Vecchio}, {Venkatraman Krishnan}, {Verbiest}, {Wang}, {Wang}, \& {Wu}}]{2023A&A...678A..48E}
{EPTA and InPTA Collaboration}, {Antoniadis}, J., {Babak}, S., {et~al.} 2023, \aap, 678, A48

\bibitem[{{EPTA and InPTA Collaborations} {et~al.}(2023){EPTA and InPTA Collaborations}, {Antoniadis}, {Arumugam}, {Arumugam}, {Babak}, {Bagchi}, {Bak Nielsen}, {Bassa}, {Bathula}, {Berthereau}, {Bonetti}, {Bortolas}, {Brook}, {Burgay}, {Caballero}, {Chalumeau}, {Champion}, {Chanlaridis}, {Chen}, {Cognard}, {Dandapat}, {Deb}, {Desai}, {Desvignes}, {Dhanda-Batra}, {Dwivedi}, {Falxa}, {Ferdman}, {Franchini}, {Gair}, {Goncharov}, {Gopakumar}, {Graikou}, {Grie{\ss}meier}, {Guillemot}, {Guo}, {Gupta}, {Hisano}, {Hu}, {Iraci}, {Izquierdo-Villalba}, {Jang}, {Jawor}, {Janssen}, {Jessner}, {Joshi}, {Kareem}, {Karuppusamy}, {Keane}, {Keith}, {Kharbanda}, {Kikunaga}, {Kolhe}, {Kramer}, {Krishnakumar}, {Lackeos}, {Lee}, {Liu}, {Liu}, {Lyne}, {McKee}, {Maan}, {Main}, {Mickaliger}, {Ni{\c{t}}u}, {Nobleson}, {Paladi}, {Parthasarathy}, {Perera}, {Perrodin}, {Petiteau}, {Porayko}, {Possenti}, {Prabu}, {Quelquejay Leclere}, {Rana}, {Samajdar}, {Sanidas}, {Sesana}, {Shaifullah}, {Singha}, {Speri}, {Spiewak}, {Srivastava},
  {Stappers}, {Surnis}, {Susarla}, {Susobhanan}, {Takahashi}, {Tarafdar}, {Theureau}, {Tiburzi}, {van der Wateren}, {Vecchio}, {Venkatraman Krishnan}, {Verbiest}, {Wang}, {Wang}, \& {Wu}}]{2023A&A...678A..50E}
{EPTA and InPTA Collaborations}, {Antoniadis}, J., {Arumugam}, P., {et~al.} 2023, \aap, 678, A50

\bibitem[{{Falxa} {et~al.}(2023){Falxa}, {Babak}, {Baker}, {B{\'e}csy}, {Chalumeau}, {Chen}, {Chen}, {Cornish}, {Guillemot}, {Hazboun}, {Mingarelli}, {Parthasarathy}, {Petiteau}, {Pol}, {Sesana}, {Spolaor}, {Taylor}, {Theureau}, {Vallisneri}, {Vigeland}, {Witt}, {Zhu}, {Antoniadis}, {Arzoumanian}, {Bailes}, {Bhat}, {Blecha}, {Brazier}, {Brook}, {Caballero}, {Cameron}, {Casey-Clyde}, {Champion}, {Charisi}, {Chatterjee}, {Cognard}, {Cordes}, {Crawford}, {Cromartie}, {Crowter}, {Dai}, {DeCesar}, {Demorest}, {Desvignes}, {Dolch}, {Drachler}, {Feng}, {Ferrara}, {Fiore}, {Fonseca}, {Garver-Daniels}, {Glaser}, {Goncharov}, {Good}, {Griessmeier}, {Guo}, {G{\"u}ltekin}, {Hobbs}, {Hu}, {Islo}, {Jang}, {Jennings}, {Johnson}, {Jones}, {Kaczmarek}, {Kaiser}, {Kaplan}, {Keith}, {Kelley}, {Kerr}, {Key}, {Laal}, {Lam}, {Lamb}, {Lazio}, {Liu}, {Liu}, {Luo}, {Lynch}, {Madison}, {Main}, {Manchester}, {McEwen}, {McKee}, {McLaughlin}, {Ng}, {Nice}, {Ocker}, {Olum}, {Os{\l}owski}, {Pennucci}, {Perera}, {Perrodin}, {Porayko},
  {Possenti}, {Quelquejay-Leclere}, {Ransom}, {Ray}, {Reardon}, {Russell}, {Samajdar}, {Sarkissian}, {Schult}, {Shaifullah}, {Shannon}, {Shapiro-Albert}, {Siemens}, {Simon}, {Siwek}, {Smith}, {Speri}, {Spiewak}, {Stairs}, {Stappers}, {Stinebring}, {Swiggum}, {Tiburzi}, {Turner}, {Vecchio}, {Verbiest}, {Wahl}, {Wang}, {Wang}, {Wang}, {Wu}, {Zhang}, {Zhang}, \& {IPTA Collaboration}}]{2023MNRAS.521.5077F}
{Falxa}, M., {Babak}, S., {Baker}, P.~T., {et~al.} 2023, \mnras, 521, 5077

\bibitem[{{Foster} \& {Backer}(1990)}]{1990ApJ...361..300F}
{Foster}, R.~S., \& {Backer}, D.~C. 1990, \apj, 361, 300

\bibitem[{{Foster} {et~al.}(1991){Foster}, {Fairhead}, \& {Backer}}]{1991ApJ...378..687F}
{Foster}, R.~S., {Fairhead}, L., \& {Backer}, D.~C. 1991, \apj, 378, 687

\bibitem[{{Gitika} {et~al.}(2023){Gitika}, {Bailes}, {Shannon}, {Reardon}, {Cameron}, {Shamohammadi}, {Miles}, {Flynn}, {Corongiu}, \& {Kramer}}]{2023MNRAS.526.3370G}
{Gitika}, P., {Bailes}, M., {Shannon}, R.~M., {et~al.} 2023, \mnras, 526, 3370

\bibitem[{{Goncharov} {et~al.}(2021){Goncharov}, {Reardon}, {Shannon}, {Zhu}, {Thrane}, {Bailes}, {Bhat}, {Dai}, {Hobbs}, {Kerr}, {Manchester}, {Os{\l}owski}, {Parthasarathy}, {Russell}, {Spiewak}, {Thyagarajan}, \& {Wang}}]{2021MNRAS.502..478G}
{Goncharov}, B., {Reardon}, D.~J., {Shannon}, R.~M., {et~al.} 2021, \mnras, 502, 478

\bibitem[{{Grishchuk}(2005)}]{2005PhyU...48.1235G}
{Grishchuk}, L.~P. 2005, Physics Uspekhi, 48, 1235

\bibitem[{Harris {et~al.}(2020)Harris, Millman, van~der Walt, Gommers, Virtanen, Cournapeau, Wieser, Taylor, Berg, Smith, Kern, Picus, Hoyer, van Kerkwijk, Brett, Haldane, del R{\'{i}}o, Wiebe, Peterson, G{\'{e}}rard-Marchant, Sheppard, Reddy, Weckesser, Abbasi, Gohlke, \& Oliphant}]{harris2020array}
Harris, C.~R., Millman, K.~J., van~der Walt, S.~J., {et~al.} 2020, Nature, 585, 357

\bibitem[{{Haslam} {et~al.}(1982){Haslam}, {Salter}, {Stoffel}, \& {Wilson}}]{1982A&AS...47....1H}
{Haslam}, C.~G.~T., {Salter}, C.~J., {Stoffel}, H., \& {Wilson}, W.~E. 1982, \aaps, 47, 1

\bibitem[{{Hazboun} {et~al.}(2019){Hazboun}, {Romano}, \& {Smith}}]{2019PhRvD.100j4028H}
{Hazboun}, J.~S., {Romano}, J.~D., \& {Smith}, T.~L. 2019, \prd, 100, 104028

\bibitem[{{Helfand} {et~al.}(1975){Helfand}, {Manchester}, \& {Taylor}}]{1975ApJ...198..661H}
{Helfand}, D.~J., {Manchester}, R.~N., \& {Taylor}, J.~H. 1975, \apj, 198, 661

\bibitem[{{Hellings} \& {Downs}(1983)}]{1983ApJ...265L..39H}
{Hellings}, R.~W., \& {Downs}, G.~S. 1983, \apjl, 265, L39

\bibitem[{{Hobbs} {et~al.}(2010{\natexlab{a}}){Hobbs}, {Lyne}, \& {Kramer}}]{2010MNRAS.402.1027H}
{Hobbs}, G., {Lyne}, A.~G., \& {Kramer}, M. 2010{\natexlab{a}}, \mnras, 402, 1027

\bibitem[{{Hobbs} {et~al.}(2010{\natexlab{b}}){Hobbs}, {Archibald}, {Arzoumanian}, {Backer}, {Bailes}, {Bhat}, {Burgay}, {Burke-Spolaor}, {Champion}, {Cognard}, {Coles}, {Cordes}, {Demorest}, {Desvignes}, {Ferdman}, {Finn}, {Freire}, {Gonzalez}, {Hessels}, {Hotan}, {Janssen}, {Jenet}, {Jessner}, {Jordan}, {Kaspi}, {Kramer}, {Kondratiev}, {Lazio}, {Lazaridis}, {Lee}, {Levin}, {Lommen}, {Lorimer}, {Lynch}, {Lyne}, {Manchester}, {McLaughlin}, {Nice}, {Oslowski}, {Pilia}, {Possenti}, {Purver}, {Ransom}, {Reynolds}, {Sanidas}, {Sarkissian}, {Sesana}, {Shannon}, {Siemens}, {Stairs}, {Stappers}, {Stinebring}, {Theureau}, {van Haasteren}, {van Straten}, {Verbiest}, {Yardley}, \& {You}}]{2010CQGra..27h4013H}
{Hobbs}, G., {Archibald}, A., {Arzoumanian}, Z., {et~al.} 2010{\natexlab{b}}, Classical and Quantum Gravity, 27, 084013

\bibitem[{Hunter(2007)}]{Hunter:2007}
Hunter, J.~D. 2007, Computing in Science \& Engineering, 9, 90

\bibitem[{{Iraci} {et~al.}(2024){Iraci}, {Chalumeau}, {Tiburzi}, {Verbiest}, {Possenti}, {Shaifullah}, {Susarla}, {Krishnakumar}, {Lam}, {Cromartie}, {Kerr}, \& {Grie{\ss}meier}}]{2024A&A...692A.170I}
{Iraci}, F., {Chalumeau}, A., {Tiburzi}, C., {et~al.} 2024, \aap, 692, A170

\bibitem[{{Jaffe} \& {Backer}(2003)}]{2003ApJ...583..616J}
{Jaffe}, A.~H., \& {Backer}, D.~C. 2003, \apj, 583, 616

\bibitem[{{Jankowski} {et~al.}(2018){Jankowski}, {van Straten}, {Keane}, {Bailes}, {Barr}, {Johnston}, \& {Kerr}}]{2018MNRAS.473.4436J}
{Jankowski}, F., {van Straten}, W., {Keane}, E.~F., {et~al.} 2018, \mnras, 473, 4436

\bibitem[{{Jenet} {et~al.}(1998){Jenet}, {Anderson}, {Kaspi}, {Prince}, \& {Unwin}}]{1998ApJ...498..365J}
{Jenet}, F.~A., {Anderson}, S.~B., {Kaspi}, V.~M., {Prince}, T.~A., \& {Unwin}, S.~C. 1998, \apj, 498, 365

\bibitem[{{Jonas} \& {MeerKAT Team}(2016)}]{2016mks..confE...1J}
{Jonas}, J., \& {MeerKAT Team}. 2016, in MeerKAT Science: On the Pathway to the SKA, 1

\bibitem[{{Kaiser} \& {McWilliams}(2021)}]{2021CQGra..38e5009K}
{Kaiser}, A.~R., \& {McWilliams}, S.~T. 2021, Classical and Quantum Gravity, 38, 055009

\bibitem[{{Kaur} {et~al.}(2022){Kaur}, {Ramesh Bhat}, {Dai}, {McSweeney}, {Shannon}, {Kudale}, \& {van Straten}}]{2022ApJ...930L..27K}
{Kaur}, D., {Ramesh Bhat}, N.~D., {Dai}, S., {et~al.} 2022, \apjl, 930, L27

\bibitem[{{Keith} {et~al.}(2013){Keith}, {Coles}, {Shannon}, {Hobbs}, {Manchester}, {Bailes}, {Bhat}, {Burke-Spolaor}, {Champion}, {Chaudhary}, {Hotan}, {Khoo}, {Kocz}, {Os{\l}owski}, {Ravi}, {Reynolds}, {Sarkissian}, {van Straten}, \& {Yardley}}]{2013MNRAS.429.2161K}
{Keith}, M.~J., {Coles}, W., {Shannon}, R.~M., {et~al.} 2013, \mnras, 429, 2161

\bibitem[{{Kerr} {et~al.}(2018){Kerr}, {Coles}, {Ward}, {Johnston}, {Tuntsov}, \& {Shannon}}]{2018MNRAS.474.4637K}
{Kerr}, M., {Coles}, W.~A., {Ward}, C.~A., {et~al.} 2018, \mnras, 474, 4637

\bibitem[{{Kramer} {et~al.}(2021){Kramer}, {Stairs}, {Venkatraman Krishnan}, {Freire}, {Abbate}, {Bailes}, {Burgay}, {Buchner}, {Champion}, {Cognard}, {Gautam}, {Geyer}, {Guillemot}, {Hu}, {Janssen}, {Lower}, {Parthasarathy}, {Possenti}, {Ransom}, {Reardon}, {Ridolfi}, {Serylak}, {Shannon}, {Spiewak}, {Theureau}, {van Straten}, {Wex}, {Oswald}, {Posselt}, {Sobey}, {Barr}, {Camilo}, {Hugo}, {Jameson}, {Johnston}, {Karastergiou}, {Keith}, \& {Os{\l}owski}}]{2021MNRAS.504.2094K}
{Kramer}, M., {Stairs}, I.~H., {Venkatraman Krishnan}, V., {et~al.} 2021, \mnras, 504, 2094

\bibitem[{{Kulkarni} {et~al.}(2024){Kulkarni}, {Shannon}, {Reardon}, {Miles}, {Bailes}, \& {Shamohammadi}}]{2024MNRAS.528.3658K}
{Kulkarni}, A.~D., {Shannon}, R.~M., {Reardon}, D.~J., {et~al.} 2024, \mnras, 528, 3658

\bibitem[{{Lam}(2018)}]{2018ApJ...868...33L}
{Lam}, M.~T. 2018, \apj, 868, 33

\bibitem[{{Lam} {et~al.}(2020){Lam}, {Lazio}, {Dolch}, {Jones}, {McLaughlin}, {Stinebring}, \& {Surnis}}]{2020ApJ...892...89L}
{Lam}, M.~T., {Lazio}, T.~J.~W., {Dolch}, T., {et~al.} 2020, \apj, 892, 89

\bibitem[{{Lam} {et~al.}(2016){Lam}, {Cordes}, {Chatterjee}, {Arzoumanian}, {Crowter}, {Demorest}, {Dolch}, {Ellis}, {Ferdman}, {Fonseca}, {Gonzalez}, {Jones}, {Jones}, {Levin}, {Madison}, {McLaughlin}, {Nice}, {Pennucci}, {Ransom}, {Siemens}, {Stairs}, {Stovall}, {Swiggum}, \& {Zhu}}]{2016ApJ...819..155L}
{Lam}, M.~T., {Cordes}, J.~M., {Chatterjee}, S., {et~al.} 2016, \apj, 819, 155

\bibitem[{{Lam} {et~al.}(2019){Lam}, {McLaughlin}, {Arzoumanian}, {Blumer}, {Brook}, {Cromartie}, {Demorest}, {DeCesar}, {Dolch}, {Ellis}, {Ferdman}, {Ferrara}, {Fonseca}, {Garver-Daniels}, {Gentile}, {Jones}, {Lorimer}, {Lynch}, {Ng}, {Nice}, {Pennucci}, {Ransom}, {Spiewak}, {Stairs}, {Stovall}, {Swiggum}, {Vigeland}, \& {Zhu}}]{2019ApJ...872..193L}
{Lam}, M.~T., {McLaughlin}, M.~A., {Arzoumanian}, Z., {et~al.} 2019, \apj, 872, 193

\bibitem[{{Lawson} {et~al.}(1987){Lawson}, {Mayer}, {Osborne}, \& {Parkinson}}]{1987MNRAS.225..307L}
{Lawson}, K.~D., {Mayer}, C.~J., {Osborne}, J.~L., \& {Parkinson}, M.~L. 1987, \mnras, 225, 307

\bibitem[{{Lazarus} {et~al.}(2016){Lazarus}, {Karuppusamy}, {Graikou}, {Caballero}, {Champion}, {Lee}, {Verbiest}, \& {Kramer}}]{2016MNRAS.458..868L}
{Lazarus}, P., {Karuppusamy}, R., {Graikou}, E., {et~al.} 2016, \mnras, 458, 868

\bibitem[{{Lee} {et~al.}(2012){Lee}, {Bassa}, {Janssen}, {Karuppusamy}, {Kramer}, {Smits}, \& {Stappers}}]{2012MNRAS.423.2642L}
{Lee}, K.~J., {Bassa}, C.~G., {Janssen}, G.~H., {et~al.} 2012, \mnras, 423, 2642

\bibitem[{{Lentati} {et~al.}(2016){Lentati}, {Shannon}, {Coles}, {Verbiest}, {van Haasteren}, {Ellis}, {Caballero}, {Manchester}, {Arzoumanian}, {Babak}, {Bassa}, {Bhat}, {Brem}, {Burgay}, {Burke-Spolaor}, {Champion}, {Chatterjee}, {Cognard}, {Cordes}, {Dai}, {Demorest}, {Desvignes}, {Dolch}, {Ferdman}, {Fonseca}, {Gair}, {Gonzalez}, {Graikou}, {Guillemot}, {Hessels}, {Hobbs}, {Janssen}, {Jones}, {Karuppusamy}, {Keith}, {Kerr}, {Kramer}, {Lam}, {Lasky}, {Lassus}, {Lazarus}, {Lazio}, {Lee}, {Levin}, {Liu}, {Lynch}, {Madison}, {McKee}, {McLaughlin}, {McWilliams}, {Mingarelli}, {Nice}, {Os{\l}owski}, {Pennucci}, {Perera}, {Perrodin}, {Petiteau}, {Possenti}, {Ransom}, {Reardon}, {Rosado}, {Sanidas}, {Sesana}, {Shaifullah}, {Siemens}, {Smits}, {Stairs}, {Stappers}, {Stinebring}, {Stovall}, {Swiggum}, {Taylor}, {Theureau}, {Tiburzi}, {Toomey}, {Vallisneri}, {van Straten}, {Vecchio}, {Wang}, {Wang}, {You}, {Zhu}, \& {Zhu}}]{2016MNRAS.458.2161L}
{Lentati}, L., {Shannon}, R.~M., {Coles}, W.~A., {et~al.} 2016, \mnras, 458, 2161

\bibitem[{{Manchester} {et~al.}(2005){Manchester}, {Hobbs}, {Teoh}, \& {Hobbs}}]{2005AJ....129.1993M}
{Manchester}, R.~N., {Hobbs}, G.~B., {Teoh}, A., \& {Hobbs}, M. 2005, \aj, 129, 1993

\bibitem[{{Middleton} {et~al.}(2025){Middleton}, {Shannon}, {Bailes}, {Cameron}, {Corongiu}, {Geyer}, {Jones}, {Kramer}, {Miles}, {Parthasarathy}, {Possenti}, \& {Reardon}}]{2025MNRAS.540..603M}
{Middleton}, H., {Shannon}, R.~M., {Bailes}, M., {et~al.} 2025, \mnras, 540, 603

\bibitem[{{Miles} {et~al.}(2022){Miles}, {Shannon}, {Bailes}, {Reardon}, {Buchner}, {Middleton}, \& {Spiewak}}]{2022MNRAS.510.5908M}
{Miles}, M.~T., {Shannon}, R.~M., {Bailes}, M., {et~al.} 2022, \mnras, 510, 5908

\bibitem[{{Miles} {et~al.}(2023){Miles}, {Shannon}, {Bailes}, {Reardon}, {Keith}, {Cameron}, {Parthasarathy}, {Shamohammadi}, {Spiewak}, {van Straten}, {Buchner}, {Camilo}, {Geyer}, {Karastergiou}, {Kramer}, {Serylak}, {Theureau}, \& {Venkatraman Krishnan}}]{2023MNRAS.519.3976M}
---. 2023, \mnras, 519, 3976

\bibitem[{{Miles} {et~al.}(2025{\natexlab{a}}){Miles}, {Shannon}, {Reardon}, {Bailes}, {Champion}, {Geyer}, {Gitika}, {Grunthal}, {Keith}, {Kramer}, {Kulkarni}, {Nathan}, {Parthasarathy}, {Porayko}, {Singha}, {Theureau}, {Abbate}, {Buchner}, {Cameron}, {Camilo}, {Moreschi}, {Shaifullah}, {Shamohammadi}, \& {Krishnan}}]{2025MNRAS.536.1467M}
{Miles}, M.~T., {Shannon}, R.~M., {Reardon}, D.~J., {et~al.} 2025{\natexlab{a}}, \mnras, 536, 1467

\bibitem[{{Miles} {et~al.}(2025{\natexlab{b}}){Miles}, {Shannon}, {Reardon}, {Bailes}, {Champion}, {Geyer}, {Gitika}, {Grunthal}, {Keith}, {Kramer}, {Kulkarni}, {Nathan}, {Parthasarathy}, {Singha}, {Theureau}, {Thrane}, {Abbate}, {Buchner}, {Cameron}, {Camilo}, {Moreschi}, {Shaifullah}, {Shamohammadi}, {Possenti}, \& {Krishnan}}]{2025MNRAS.536.1489M}
---. 2025{\natexlab{b}}, \mnras, 536, 1489

\bibitem[{{NANOGrav Collaboration} {et~al.}(2015){NANOGrav Collaboration}, {Arzoumanian}, {Brazier}, {Burke-Spolaor}, {Chamberlin}, {Chatterjee}, {Christy}, {Cordes}, {Cornish}, {Crowter}, {Demorest}, {Dolch}, {Ellis}, {Ferdman}, {Fonseca}, {Garver-Daniels}, {Gonzalez}, {Jenet}, {Jones}, {Jones}, {Kaspi}, {Koop}, {Lam}, {Lazio}, {Levin}, {Lommen}, {Lorimer}, {Luo}, {Lynch}, {Madison}, {McLaughlin}, {McWilliams}, {Nice}, {Palliyaguru}, {Pennucci}, {Ransom}, {Siemens}, {Stairs}, {Stinebring}, {Stovall}, {Swiggum}, {Vallisneri}, {van Haasteren}, {Wang}, \& {Zhu}}]{2015ApJ...813...65N}
{NANOGrav Collaboration}, {Arzoumanian}, Z., {Brazier}, A., {et~al.} 2015, \apj, 813, 65

\bibitem[{{Natarajan} {et~al.}(2024){Natarajan}, {Pacucci}, {Ricarte}, {Bogd{\'a}n}, {Goulding}, \& {Cappelluti}}]{2024ApJ...960L...1N}
{Natarajan}, P., {Pacucci}, F., {Ricarte}, A., {et~al.} 2024, \apjl, 960, L1

\bibitem[{{{\"O}lmez} {et~al.}(2010){{\"O}lmez}, {Mandic}, \& {Siemens}}]{2010PhRvD..81j4028O}
{{\"O}lmez}, S., {Mandic}, V., \& {Siemens}, X. 2010, \prd, 81, 104028

\bibitem[{{Parthasarathy} {et~al.}(2021){Parthasarathy}, {Bailes}, {Shannon}, {van Straten}, {Os{\l}owski}, {Johnston}, {Spiewak}, {Reardon}, {Kramer}, {Venkatraman Krishnan}, {Pennucci}, {Abbate}, {Buchner}, {Camilo}, {Champion}, {Geyer}, {Hugo}, {Jameson}, {Karastergiou}, {Keith}, \& {Serylak}}]{2021MNRAS.502..407P}
{Parthasarathy}, A., {Bailes}, M., {Shannon}, R.~M., {et~al.} 2021, \mnras, 502, 407

\bibitem[{{Pennucci}(2019)}]{2019ApJ...871...34P}
{Pennucci}, T.~T. 2019, \apj, 871, 34

\bibitem[{{Pennucci} {et~al.}(2016){Pennucci}, {Demorest}, \& {Ransom}}]{2016ascl.soft06013P}
{Pennucci}, T.~T., {Demorest}, P.~B., \& {Ransom}, S.~M. 2016, {Pulse Portraiture: Pulsar timing}, Astrophysics Source Code Library, record ascl:1606.013

\bibitem[{{Reardon} {et~al.}(2023{\natexlab{a}}){Reardon}, {Zic}, {Shannon}, {Hobbs}, {Bailes}, {Di Marco}, {Kapur}, {Rogers}, {Thrane}, {Askew}, {Bhat}, {Cameron}, {Cury{\l}o}, {Coles}, {Dai}, {Goncharov}, {Kerr}, {Kulkarni}, {Levin}, {Lower}, {Manchester}, {Mandow}, {Miles}, {Nathan}, {Os{\l}owski}, {Russell}, {Spiewak}, {Zhang}, \& {Zhu}}]{2023ApJ...951L...6R}
{Reardon}, D.~J., {Zic}, A., {Shannon}, R.~M., {et~al.} 2023{\natexlab{a}}, \apjl, 951, L6

\bibitem[{{Reardon} {et~al.}(2023{\natexlab{b}}){Reardon}, {Zic}, {Shannon}, {Di Marco}, {Hobbs}, {Kapur}, {Lower}, {Mandow}, {Middleton}, {Miles}, {Rogers}, {Askew}, {Bailes}, {Bhat}, {Cameron}, {Kerr}, {Kulkarni}, {Manchester}, {Nathan}, {Russell}, {Os{\l}owski}, \& {Zhu}}]{2023ApJ...951L...7R}
---. 2023{\natexlab{b}}, \apjl, 951, L7

\bibitem[{{Romano} \& {Allen}(2024)}]{2024CQGra..41q5008R}
{Romano}, J.~D., \& {Allen}, B. 2024, Classical and Quantum Gravity, 41, 175008

\bibitem[{{Sesana}(2013)}]{2013MNRAS.433L...1S}
{Sesana}, A. 2013, \mnras, 433, L1

\bibitem[{{Sesana} {et~al.}(2008){Sesana}, {Vecchio}, \& {Colacino}}]{2008MNRAS.390..192S}
{Sesana}, A., {Vecchio}, A., \& {Colacino}, C.~N. 2008, \mnras, 390, 192

\bibitem[{{Sesana} {et~al.}(2009){Sesana}, {Vecchio}, \& {Volonteri}}]{2009MNRAS.394.2255S}
{Sesana}, A., {Vecchio}, A., \& {Volonteri}, M. 2009, \mnras, 394, 2255

\bibitem[{{Shamohammadi} {et~al.}(2024){Shamohammadi}, {Bailes}, {Flynn}, {Reardon}, {Shannon}, {Buchner}, {Cameron}, {Camilo}, {Corongiu}, {Geyer}, {Kramer}, {Miles}, \& {Spiewak}}]{2024MNRAS.530..287S}
{Shamohammadi}, M., {Bailes}, M., {Flynn}, C., {et~al.} 2024, \mnras, 530, 287

\bibitem[{{Shannon} \& {Cordes}(2010)}]{2010ApJ...725.1607S}
{Shannon}, R.~M., \& {Cordes}, J.~M. 2010, \apj, 725, 1607

\bibitem[{{Shannon} \& {Cordes}(2017)}]{2017MNRAS.464.2075S}
---. 2017, \mnras, 464, 2075

\bibitem[{{Shannon} {et~al.}(2014){Shannon}, {Os{\l}owski}, {Dai}, {Bailes}, {Hobbs}, {Manchester}, {van Straten}, {Raithel}, {Ravi}, {Toomey}, {Bhat}, {Burke-Spolaor}, {Coles}, {Keith}, {Kerr}, {Levin}, {Sarkissian}, {Wang}, {Wen}, \& {Zhu}}]{2014MNRAS.443.1463S}
{Shannon}, R.~M., {Os{\l}owski}, S., {Dai}, S., {et~al.} 2014, \mnras, 443, 1463

\bibitem[{{Sieber} {et~al.}(1975){Sieber}, {Reinecke}, \& {Wielebinski}}]{1975A&A....38..169S}
{Sieber}, W., {Reinecke}, R., \& {Wielebinski}, R. 1975, \aap, 38, 169

\bibitem[{{Siemens} {et~al.}(2013){Siemens}, {Ellis}, {Jenet}, \& {Romano}}]{2013CQGra..30v4015S}
{Siemens}, X., {Ellis}, J., {Jenet}, F., \& {Romano}, J.~D. 2013, Classical and Quantum Gravity, 30, 224015

\bibitem[{{Slee} {et~al.}(1987){Slee}, {Bobra}, \& {Alurkar}}]{1987AuJPh..40..557S}
{Slee}, O.~B., {Bobra}, A.~D., \& {Alurkar}, S.~K. 1987, Australian Journal of Physics, 40, 557

\bibitem[{{Speri} {et~al.}(2023){Speri}, {Porayko}, {Falxa}, {Chen}, {Gair}, {Sesana}, \& {Taylor}}]{2023MNRAS.518.1802S}
{Speri}, L., {Porayko}, N.~K., {Falxa}, M., {et~al.} 2023, \mnras, 518, 1802

\bibitem[{{Spiewak} {et~al.}(2022){Spiewak}, {Bailes}, {Miles}, {Parthasarathy}, {Reardon}, {Shamohammadi}, {Shannon}, {Bhat}, {Buchner}, {Cameron}, {Camilo}, {Geyer}, {Johnston}, {Karastergiou}, {Keith}, {Kramer}, {Serylak}, {van Straten}, {Theureau}, \& {Venkatraman Krishnan}}]{2022PASA...39...27S}
{Spiewak}, R., {Bailes}, M., {Miles}, M.~T., {et~al.} 2022, \pasa, 39, e027

\bibitem[{{Vallisneri}(2020)}]{2020ascl.soft02017V}
{Vallisneri}, M. 2020, {libstempo: Python wrapper for Tempo2}, Astrophysics Source Code Library, record ascl:2002.017

\bibitem[{{van Haasteren} \& {Vallisneri}(2014)}]{2014PhRvD..90j4012V}
{van Haasteren}, R., \& {Vallisneri}, M. 2014, \prd, 90, 104012

\bibitem[{{Wang} {et~al.}(2024){Wang}, {Wang}, {Wang}, {Hobbs}, {Xu}, {Wang}, {Dai}, {Dang}, {Li}, {Feng}, \& {Zhang}}]{2024ApJ...964....6W}
{Wang}, S.~Q., {Wang}, N., {Wang}, J.~B., {et~al.} 2024, \apj, 964, 6

\bibitem[{{Wyithe} \& {Loeb}(2003)}]{2003ApJ...595..614W}
{Wyithe}, J. S.~B., \& {Loeb}, A. 2003, \apj, 595, 614

\bibitem[{{Xu} {et~al.}(2023){Xu}, {Chen}, {Guo}, {Jiang}, {Wang}, {Xu}, {Xue}, {Nicolas Caballero}, {Yuan}, {Xu}, {Wang}, {Hao}, {Luo}, {Lee}, {Han}, {Jiang}, {Shen}, {Wang}, {Wang}, {Xu}, {Wu}, {Manchester}, {Qian}, {Guan}, {Huang}, {Sun}, \& {Zhu}}]{2023RAA....23g5024X}
{Xu}, H., {Chen}, S., {Guo}, Y., {et~al.} 2023, Research in Astronomy and Astrophysics, 23, 075024

\bibitem[{{Zic} {et~al.}(2023){Zic}, {Reardon}, {Kapur}, {Hobbs}, {Mandow}, {Cury{\l}o}, {Shannon}, {Askew}, {Bailes}, {Bhat}, {Cameron}, {Chen}, {Dai}, {Di Marco}, {Feng}, {Kerr}, {Kulkarni}, {Lower}, {Luo}, {Manchester}, {Miles}, {Nathan}, {Os{\l}owski}, {Rogers}, {Russell}, {Sarkissian}, {Shamohammadi}, {Spiewak}, {Thyagarajan}, {Toomey}, {Wang}, {Zhang}, {Zhang}, \& {Zhu}}]{2023PASA...40...49Z}
{Zic}, A., {Reardon}, D.~J., {Kapur}, A., {et~al.} 2023, \pasa, 40, e049

\end{thebibliography}



\end{document}